\documentclass[fleqn,usenatbib]{mnras}
\usepackage{amsmath,amssymb}
\usepackage{txfonts}
\usepackage[T1]{fontenc}
\usepackage{ae,aecompl}

\usepackage{tabularx,graphicx,multirow}

\newcommand{\plotpath}[1]{#1} 
\newcommand{\eff}{\text{eff}} 
\newcommand{\flux}{F} 
\newcommand{\ADU}{\textrm{ADU}} 
\newcommand{\cald}[1]{#1^\textrm{c}} 
\newcommand{\true}[1]{#1^\star} 
\newcommand{\meadf}[1]{#1^\ADU} 
\newcommand{\mead}[1]{#1} 
\newcommand{\exps}{\textrm{exp}} 
\newcommand{\overlap}{\textrm{o}} 
\newcommand{\abs}[1]{\lvert#1\rvert} 

\title{Large-scale Retrospective Relative Spectro-photometric Self-calibration in Space}
\author[Markovi\v{c} et al.]{\parbox{\textwidth}{\Large%
Katarina Markovi\v{c},$^{1}$\thanks{E-mail: dida.markovic@port.ac.uk (KM)}
Will J. Percival,$^{1}$
Marco Scodeggio,$^{2}$
Anne Ealet,$^{3}$
Stefanie Wachter,$^{4}$
Bianca Garilli,$^{2}$
Luigi Guzzo,$^{5,6}$
Roberto Scaramella,$^{7}$
Elisabetta Maiorano,$^{8}$
and J\'er\^ome Amiaux$^{9}$
}\vspace*{4pt} \\ 
$^{1}$Institute of Cosmology and Gravitation, University of Portsmouth, Dennis Sciama Building, Portsmouth P01 3FX, UK\\
$^{2}$INAF-IASF, Via Bassini 15, I-20133 Milano, Italy\\
$^{3}$Centre de Physique des Particules de Marseille, Aix-Marseille Universit\'e, CNRS/IN2P3, F-13288 Marseille, France\\
$^{4}$Max Planck Institute for Astronomy, K\"onigstuhl 17, D-69117 Heidelberg, Germany\\
$^{5}$Dipartimento di Fisica, Universit\`a degli Studi di Milano, via G. Celoria 16, 20133 Milano, Italy\\
$^{6}$INAF-Osservatorio Astronomico di Brera, Via E. Bianchi 46, I-23807 Merate, Italy\\
$^{7}$INAF-Osservatorio di Roma, Via Frascati 33, I-00040 Monte Porzio Catone (Roma), Italy\\
$^{8}$INAF-IASF Bologna, via Piero Gobetti 101, 40129 Bologna, Italy\\
$^{9}$Commissariat \`a l'\'Energie Atomique, Orme des Merisiers, F-91191 Gif sur Yvette, France
}

\date{Accepted 2017 January 31. Received 2017 January 30; in original form 2016 June 22}

\pubyear{2016}

\begin{document}
\label{firstpage}
\pagerange{\pageref{firstpage}--\pageref{lastpage}}
\maketitle

\begin{abstract}
We consider the application of relative self-calibration using overlap regions to spectroscopic galaxy surveys that use slit-less spectroscopy. This method is based on that developed for the SDSS by \citet{Padmanabhan:2008} in that we consider jointly fitting and marginalising over calibrator brightness, rather than treating these as free parameters. However, we separate the calibration of the detector-to-detector from the full-focal-plane exposure-to-exposure calibration. To demonstrate how the calibration procedure will work, we simulate the procedure for a potential implementation of the spectroscopic component of the wide Euclid survey. We study the change of coverage and the determination of relative multiplicative errors in flux measurements for different dithering configurations. We use the new method to study the case where the flat-field across each exposure or detector is measured precisely and only exposure-to-exposure or detector-to-detector variation in the flux error remains. We consider several base dither patterns and find that they strongly influence the ability to calibrate, using this methodology. To enable self-calibration, it is important that the survey strategy connects different observations with at least a minimum amount of overlap, and we propose an ``S''-pattern for dithering that fulfills this requirement. 
The final survey strategy adopted by Euclid will have to optimise for a number of different science goals and requirements. The large-scale calibration of the spectroscopic galaxy survey is clearly cosmologically crucial, but is not the only one.\\
We make our simulation code public on \href{http://github.com/didamarkovic/ubercal}{\texttt{github.com/didamarkovic/ubercal}}.
\end{abstract}


\section{Introduction}

The analysis of intrinsic galaxy clustering inherently relies on accurate sample density calibration. Any fluctuations in the density of galaxies caused by mis-calibration have the potential to be misinterpreted as being of astrophysical origin, which would bias measurements. For cosmological applications it is large-scale relative calibration that is important, as the signal lies in the fractional variation of galaxy density, rather than the absolute value. It is also not the calibration of individual galaxies that is important, but large-scale variations of the average calibration. Such relative calibration has been shown to be a dominant systematic in current large-scale clustering measurements from the Sloan Digital Sky Survey (SDSS) \citep{Leistedt:2015,Agarwal:2014,Vargas:2013, Ross:2012,Ho:2012,Ross:2011} and the Dark Energy Survey (DES) \citep{Elsner:2015, Crocce:2015, Leistedt:2015}.

For galaxy surveys using photometric redshifts, or spectroscopic samples where targets are selected from imaging data, any angular mis-calibration of the photometry directly translates into spatial fluctuations in the resulting spectroscopic catalogue. For spectroscopic surveys, we also have to calibrate observation-dependent variations in redshift measurement efficiency. The CMASS sample from the Baryon Oscillation Spectroscopic Survey \citep{Dawson:2013}, part of SDSS-III \citep{Eisenstein:2011}, has an average classification completeness of 96\%. The high completeness means that variations in spectral signal-to-noise only have a small effect on the measured catalogue, as most redshift measurements are secure even at lower signal-to-noise. Future surveys, which propose to make spectral measurements that have emission line signal-to-noise that is closer to the limit required for redshift measurement, will be more sensitive to spectrophotometric calibration, and have lower overall completeness. i.e. the calibration is sensitive to the gradient $dn_{\rm gal}/d{\rm SNR}$ at the emission line signal to noise (SNR) required for redshift measurement. Therefore the local flux limits in the survey, or equivalently, the local magnitude zero-points must be determined and corrected. This is the process of calibration.

Euclid \citep{Laureijs:2011,Amendola:2013} is the second medium-size (M-class) mission in the ESA cosmic visions program and will cover two primary probes of cosmic large-scale structure: weak lensing and gravitational clustering. Based on these two probes, Euclid will map out the underlying distribution of dark matter in the universe, as well as constrain the cosmic expansion history and therefore the equation of state parameter of dark energy. For each probe the telescope will simultaneously investigate, a separate instrument is being built: the VISual (VIS) instrument \citep{Cropper:2014} for imaging in the optical, and the Near-Infrared Spectroscope and Photometer (NISP) instrument \citep[][]{Maciaszek:2014} for imaging and slit-less spectroscopy in the near infra-red (NIR). In order to achieve its scientific goals, the Euclid mission must be optimised  in a way that balances the requirements of these two primary probes, with additional science goals. 

Euclid will undertake a slit-less spectroscopic galaxy survey using a NIR grism in the light path to decompose incident light. A detection in the NIR-imaging data will be used to locate the galaxies and provide the baseline for the wavelength solution for the NIR slit-less spectroscopy. The response of the grism is designed to enable the detection and wavelength measurement of the H-$\alpha$ line emitted by galaxies with redshifts $0.9<z<1.8$. 
Whereas strong fluctuations in the calibration of the NIR imaging data may scatter galaxies back across this detection threshold, we do not expect many galaxies for which we will have H-$\alpha$ detections failing to be detected in a stacked NIR-image. Therefore the dominant dependence will be on the spectrophotometric calibration of the grism observations.
 
A review of the systematic uncertainties that need to be removed from ground-based CCD surveys is presented in \citet{Stubbs:2006} and references therein, which also compares calibration procedures. In general, the calibration of measurements can proceed in a number of ways. The traditional method for calibration is to match observations to standards with known brightnesses, which is good at fixing the absolute calibration of individual objects. However, in order to calibrate a survey using such a methodology, many observations of standards are required, with the observing frequency dependent on the rate of drift in the measurement response. An alternative is self-calibration, where measurements of the same object, taken using different overlapping exposures within a single survey are used to limit variations between exposures \citep[e.g.][]{Padmanabhan:2008}. Such methods are good at removing small-scale variations, but can have limited sensitivity to large-scale drifts in calibration, which are spread into a coherent small offsets in a large number of overlaps. This is a function of survey strategy. Thus, regular observations of standards are still required, but with lower frequency, to correct the large-scale drifts. Recently, \citet{Finkbeiner:2015, Shafer:2015, Holmes:2012, Schlafly:2012} employed and expanded upon the approach of \citet{Padmanabhan:2008}, looking at the impact on survey design and possible other extensions of the procedure. The procedure has been applied in the cases of two ground-based surveys: SDSS and PanSTARRS-1 \citep[PS1,][]{Kaiser:2010}. However, Euclid is a space-based step-and-stare survey. Therefore its calibration will not use large, interwoven, contiguous scans to connect the different survey regions and epochs. This will be achieved by returning to the same field(s) to monitor the long-term instrument stability on large space- and time-scales and with small overlaps of adjacent fields to ensure the stability between those visits.

The overlap-based \textit{ubercalibration} procedure developed by \cite{Padmanabhan:2008} was applied to the SDSS imaging data and provided relative photometry of the order of 1\%. The procedure performs a likelihood maximisation, jointly fitting for the parameters of their calibration model, having marginalised over the magnitudes of stars in overlap regions. The calibration model was designed to combine knowledge of the frequency of different potential calibration changes and their form, with free parameters to quantify the amplitudes. Here, we consider a similar approach for calibrating the relative spectrophotometric component of the Euclid survey, for a set of different observing strategy configurations. For this component, we do not yet have a good model for the expected variations in detector response, and so we consider two models where the magnitude zero-point calibration is independently determined for each full focal plane exposure, or for each detector in each exposure. We assume that the absolute calibration will be done independently. The absolute calibration can be considered to be on the largest scale - that of the entire survey, but this is not discussed in this paper. Assuming the absolute calibration has determined the total zero-point of the survey, the remaining non-uniformity of the survey zero-points can be mitigated with relative calibration, which is our subject of discussion.

We further split the relative calibration into three types of scales: large, intermediate and small. The small scales are essentially only the pixel-to-pixel flat-field uncertainties as considered in \citet{Holmes:2012}. The intermediate, considered in this paper are the variations in the zero points of the individual exposures or at least detector tiles of the survey. We leave the consideration of the very largest scale, long term stability of the calibration to future work.

This paper presents a study on the optimisation of the observing strategy for the isolated purpose of spectro-photometric calibration. 
We describe an overlap-based model for relative calibration set up to test the impact of the observing strategy on this type of calibration. 
We have based our model observing strategies on some of the expected Euclid configuration parameter values, but we also made several simplifications, since we are only interested in the relative results. The final Euclid strategy has to take into account the visible channel as well as other limitation on survey strategy to optimise joint measurements. Indeed, in this paper we consider parts of the parameter space that are presently forbidden for the full Euclid case. However, this is justified since we expect that parts of this space may become accessible in the future due to developments during the Euclid optimisation phase.

We construct a simplified model of the Euclid spectroscopic survey with a random set of zero-point errors and describe the simulation procedure in \autoref{sec:sims}. In \autoref{sec:likeli} we calculate the posterior distribution of the calibration parameters and show the simplifications we make to find the optimal solution. In \autoref{sec:Euclid} we describe the spectroscopic component of the Euclid survey and our simplified model of it. We also describe the considered dither strategy and instrumental setup. We then attempt to perform an \textit{ubercalibration} of our synthetic survey by finding a new, corrected set of zero-points by minimizing a $\chi^2$-like mis-calibration cost function. In \autoref{sec:results} we present the results of our simulations of the relative self-calibration procedure. We summarise and conclude in \autoref{sec:conc}.

\section{Simulation Procedure}\label{sec:sims}

We would like to simulate a realistic yet simplified synthetic survey with a set of offsets of exposure zero-points to be calibrated, one for each exposure taken.

To do this, we firstly take an input survey configuration and use the \href{https://github.com/mollyswanson/mangle}{Mangle code} \citep{Swanson:2008} to create the survey geometry by inputting a list of rectangles  in the form of Mangle polygons, representing the overlapping sets of exposures of the individual NIR detectors. We then balkanise them into unique, non-overlapping regions, which we call \textbf{overlap tiles}, while keeping track of the set of detectors that cover each of them. We then treat these overlap tiles separately. We discard any tiles that are covered with one or no passes. We describe the implementation of this geometry further in \autoref{sec:Euclid}.

Secondly, we create the synthetic sky divided into overlap tiles, from the information about stellar calibrator populations. For the case of our simplified model of the spectroscopic component of Euclid, the calibrator populations are described in \autoref{sec:stars}. We employ the simplification of only generating average magnitudes optimally weighted over all the calibrator populations in each overlap tile, so that we may simulate only one value and its uncertainty per overlap, thereby speeding up the calculation of the calibration fit. This is described in more detail below in \autoref{sec:patches}.

\textit{Nota bene}, when creating the synthetic calibrator stars for the use in our simplified ubercalibration we never generate individual dispersed spectra. Our analysis starts at the point where spectra - or parts of them - have been integrated over, the fluxes converted to magnitudes, and the magnitudes averaged over the overlap tile. For this, we neglect edge effects caused by the different dispersion directions, and assume that we can calibrate using any part of the spectrum that falls into the overlap region with an accuracy depending on the amount of that spectrum in the overlap region. In other words, we are effectively generating weighted mean magnitude of all the pixels in an exposure that we expect to belong to stars. Hence, the number of calibrators that can be used can be approximated by the number of calibrators with coordinates in the overlap region.\\
Indeed, working with the full spectrum may enable us in the future to consider a chromatic dependence of the calibration and therefore enable the study of the color-dependent stability. We leave this to future work.

We add the simulated average stellar calibrator fluxes to a \textit{randomly distributed} set of initial calibrations or initial zero-point errors of the full field-of-view exposures. The initial distribution is assumed to have a width of $\sigma^\textrm{k} = 40$ millimagnitudes (mmag) in $J_\textrm{AB}$ magnitude, which corresponds to a 4\% scatter between the measured flux of full field-of-view exposures (as we show below in \autoref{eq:scatter}). \\
It is important to note that such an initial scenario of random and independent initial zero-point scatter is unrealistic and should be considered as the most significant approximation of this paper.

Finally, we perform our calibration procedure, which we describe in more detail in \autoref{sec:likeli} and find the standard deviation of the best-fit set of exposure calibration parameters $\{k_i\}$, which we call the final scatter, $\sigma^\textrm{k}_\textrm{f}$. We use this as a quality metric for the final uniformity of the survey (i.e. decrease in the scatter of the residual zero-point) after our procedure. Note that the $\sigma$ here represent the scatter around the survey mean, since we'd like to exclude any absolute calibration error from our consideration. We are only concerned with the resulting survey uniformity.

For simplicity and clarity we first simulate the calibration of entire exposures, taking the zero-point to be uniform over the focal-plane, not varying among detectors. In the second step we consider the relative calibration of individual detectors to see how much our resulting calibration scatter degrades. We report on the results in \autoref{sec:results}.

\subsection{The Spectrophotometric Model}\label{sec:model}

We now describe our simplified approach to modelling the incident calibrator brightness, the error on its measurement and the zero-point off-set in the measurement before relative calibration. We simulate both the measurement errors (\autoref{eq:notindep}) and the zero-point off-sets of individual full-focal plane exposures (\autoref{eq:scatter}) below by generating them from their respective Gaussian distributions. We justify this by first discussing the model in terms of the spectrophotometric flux, which is relevant particularly in our case of slit-less spectroscopy measurement. We convert the fluxes and their scatter to magnitudes as is also done in the case of pure photometric measurements by \citet{Padmanabhan:2008}.

\subsubsection{In terms of incident calibrator flux}

Ordinarily, we would like to make several measurements\footnote{ADU stands for Analogue-to-Digital units in which flux is measured, otherwise also called ``counts''.}, $\meadf{\flux_{ij}}$ of the fluxes of our calibrator stars, $\true{\flux}_j$. We would like to compare these measurements between the different exposures, $i$ to calibrate the exposures, optimising for this task rather than for the optimal measurement of the true fluxes. 
Following \citet{Padmanabhan:2008}, we construct the following measured, $\meadf{\flux_{ij}}$, and calibrated, $\cald{\flux_{ij}}$, quantities:
\begin{equation}
\cald{\flux_{ij}} = K_i \left(\meadf{\flux_{ij}} - \flux^0_i\right) \ ,
\label{eq:flux}
\end{equation}
where $K_i$ is the multiplicative calibration and $K_i\flux^0_i$ is the total dark current of exposure $i$. Here we assume that there is no variation in the calibration across the detectors i.e. 
\begin{equation}
K_i(x,y) = K_i + G_i(x,y) = K_i \ \textrm{with}\  G_i(x,y)=0\ \forall (x,y)\ ,
\end{equation}
where $x$ and $y$ are the pixel coordinates within the detector image of exposure $i$ \citep[such variation was studied by][we briefly revisit their method with an updated configuration in \autoref{app:FF} in the Appendix]{Holmes:2012}. This assumption is a reasonable approximation at this stage as we expect that the pixel-to-pixel flat-fields will be constrained otherwise and that the fluctuations of fluxes within the detector are strongly dominated by the fluctuations in the true fluxes of the calibrator objects. Therefore we leave the impact of the (possibly time-dependent) flat-field uncertainty on the calibration to future study.

For self-calibration, we are only interested in obtaining the $K_i$ from \autoref{eq:flux} and regard the true stellar fluxes, $\true{F_j}$ as nuisance parameters. We assume the dark current to be successfully subtracted prior to calibration, so we let $\flux^0_i \rightarrow 0$ in \autoref{eq:flux}. We then take the $-2.5\log_{10}$ of \autoref{eq:flux} (cancelling the reference object flux) and work with magnitudes from now on.

\subsubsection{In terms of incident calibrator magnitude}

The relative flux calibration, $K_i$ now becomes an additive constant, $k_i$:
\begin{equation}
\cald{m_{ij}} = \mead{m}_{ij} + k_i\ . \label{eq:mags}
\end{equation}

Before moving on to the likelihood, let us consider the distribution of calibrations. We can show that the width of the assumed Gaussian distribution, $\sigma^\text{k}$ of the exposure calibrations in magnitude, $k_i$ relates to the width, $\Sigma^\text{K}_j/F_j^\star$ of the assumed lognormal distribution of absolute calibrations residuals in flux (magnitude zero-points), $K_i$:
\begin{equation}
\sigma^\textrm{k}
= \frac{\Sigma^\textrm{K}_j}{F_j^\star} \frac{2.5}{\left<K_i\right>\ln(10)} \approx \frac{\Sigma^\textrm{K}_j}{F_j^\star} \ , \label{eq:scatter}
\end{equation}
since $\left<K_i\right> = 1$ can be assumed by requiring the absolute calibration to have been corrected perfectly beforehand by subtracting the average flux over the whole survey (see also \autoref{app:dist} in the Appendix). In words, this means that the scatter in the fractional flux variation from calibration equals the variation of additive magnitude calibrations.
Note that the $K_i$ and therefore the $k_i$ are not measurement errors, but rather the set of parameters to be determined. Their distribution should therefore be used as a prior. \\

We now also assume\footnote{Although this assumption does not impact our results, since we average over the calibrators later on, in \autoref{sec:sims}. Therefore, the Central Limit Theorem will apply largely, making the distribution of our final variables Gaussian or nearly so.} that the measurements are distributed as according to a Gaussian distribution around the true value of the magnitude.
We can construct a multivariate likelihood from the assumed probability distribution of residual ``miscalibrations'', $e_{ij}$.
The likelihood will reach a maximum when the measured and calibrated magnitude $\cald{m}_{ij}$ is precisely correct, i.e. equal to the true stellar magnitude $\true{m_j}$:
\begin{equation}
e_{ij} = \cald{m}_{ij} - \true{m_j} = 0
\label{eq:maximum}
\end{equation}
and have the width $\sigma_{ij}$ for each of the stars, $j$ and exposures $i$. The number of dimensions of the multivariate distribution will equal the number of stars plus the number of exposures, $N_\star + N_\exps$.

One may be inclined to assume that i) the true stellar brightnesses are completely independent, ii) that there is no cross-covariance between the stars and exposures and iii) that the exposure calibrations are independent from each other. The easiest to immediately justify is assumption i). Assumption ii) on the other hand can be understood as writing the variance of the $e_{ij}$ as:
\begin{equation}
\sigma_{ij}^2 = \frac{\left(\sigma_i^{\textrm{e}^-}\right)^2}{N^\textrm{pix}_j} \ ,
\label{eq:notindep}
\end{equation}
where the $\sigma_i^{\textrm{e}^-}(x, y)$ is the measurement scatter in a pixel, at $(x,y)$ in the focal plane within exposure $i$. The $(x,y)$-dependence is excluded from our analysis, because this flat-field calibration will be done using other calibration techniques as mentioned above, therefore $\sigma_i^{\textrm{e}^-}(x, y) = \sigma_i^{\textrm{e}^-}$. 
It is important to note that the dependence of the exposure-to-exposure variance on the star index, $j$ is artificial due to selecting calibrator star image pixels from the full exposure. Below, in \autoref{sec:patches}, we easily integrate out the $N_j^\textrm{pix}$-dependence, by averaging over all the calibrators in an overlap-tile. \\
Finally, in order to assume independence of adjacent exposures (iii), we must assume that the model of the decay in the detector response is either negligible or effectively fully understood. The first is unlikely to be the case, which in fact brings about the motivation for self-calibration and this paper. We adopt the second assumption for now, and leave the discussion to further work as it requires better models of long term behaviour of the instrument, but we stress here, that this stands to introduce a strong limit on the capability of the relative calibration procedure. 

Note that our notation is such that flux-related variables are capitalised and those in lower case are related to magnitudes.
We note that our approach differs somewhat from that of \citet{Padmanabhan:2008}. Firstly, the spectrophotometric model we are using is simpler, because we are considering a space-based survey, where the complex atmospheric effects do not apply. Secondly, we are considering different detectors. Whereas SDSS imaging used Charge-Coupled Devices (CCDs), Euclid spectroscopy will be using H2RG NIR detectors \citep{Beletic:2008}. 
This means that the noise model will differ. However this is not relevant for the present modelling due to the simplifications we make. The final difference is our simplification, which optimises for the measurement of the exposure calibrations and averages out any information about the individual fluxes of the calibrator stars.

Throughout the paper we use 4 types of indices: $j$ always indexes stars, $i$ exposures, $l$ overlap tiles and $p$ stellar types or ``populations''.

\subsection{Simplified Simulation of Calibrators}\label{sec:patches}

We propose to simplify the simulation of calibrator stars needed for the ubercalibration procedure. Instead of generating the individual stars (indexed with $j$) from a known distribution (see \autoref{sec:stars}), we produce a single, weighted measurement for a contiguous region of the field of view (FOV), where we assume the calibration value will be uniform. Ultimately the shape and size of such a region depends on our division of the calibration into different scales. A sensible choice for such a regions is what we call an overlap tile or overlap region. Such overlap tiles can be seen as different shaded rectangles (depending on the level of coverage) in \autoref{fig:patterns}.

The full focal plane exposures are fragmented into detector-squares and detector-gap-rectangles, which are further divided into contiguous rectangles, each covered with the same set of exposures. We refer to these latter rectangles as overlap tiles and denote them with the index $l$.
In such small patches it can be assumed that the total stellar surface number density, $n^\star(\alpha, \delta) \sim n^\star(b) = $ const for $b=80^\circ$, which is a conservative choice in calibrator density.

Therefore we end up with a measured and calibrated set of $l$ tile means in exposure $i$:
\begin{equation}
\cald{m}_{il} = \frac{1}{N^\star_l}\sum_j^{N^\star_l} \cald{m}_{ij}
\label{eq:meanf}
\end{equation}
where $N^\star_l$ is the number of all stars with $13.5~<~J_\text{AB}~<~20.5$ in the $l^\text{th}$ overlap tile and each $\cald{m}_{il}$ is averaged over all stars, $\{j\}$ in an overlap tile, $l$. 

In practice, we do not simulate the individual stars in order to average over them, but find the $\cald{m}_{il}$ from a weighted mean over the stellar populations in different brightness bins (as we do in \autoref{sec:stars}). Assuming the magnitude scatter depends only on the magnitude bin, $p$,
\begin{equation}
\sigma_p = \left(\sigma_p^{\textrm{e}^-}\right)^2 \bigg/ \sqrt{N^\textrm{pix}_j}
\approx \text{RMS}_p = \Sigma_\text{int} \bigg/ \true{F}_\text{int} \ ,
\label{eq:sigmap}
\end{equation}
we drop the index $j$, assuming $N^\textrm{pix}_j$ to be the same for all the magnitude bins, in \autoref{eq:notindep} and 
divide the sum in \autoref{eq:meanf} into $p$ sub-sums, each time summing over all the $p$-type stars in the area of the $l$-overlap, $N_{pl}^\star$. We use the RMS$_p$ values of type-F stars in \autoref{tab:snrs} to be conservative as these have the largest scatter. We obtain the $N_{pl}^\star$ each time by drawing from a Poisson distribution with mean~$=~n_p^\star A_l$, having calculated the overlap-tile-area, $A_l$ from the survey geometry and the $n_p^\star$ from \autoref{tab:Nstars} below. The $\true{\sigma_p}$ is the intrinsic, true scatter in magnitudes of $p$-type stars. \\
Note that such a division into populations is only done to simplify the simulation. When it comes to a real measurement of the stellar magnitudes, this weighting can be done with any other set of errors, which can be specified as it is deemed practical, which includes having individual weights for all the individual stars. In order to self-calibrate the survey using real data in the future, the full set of calibrator stars, including variable stars and other outliers will have to be considered.\\
Then the mean measured flux of all the stars of brightness $p$ in exposure $i$ of overlap tile $l$ would be:
\begin{equation}
\cald{m}_{ilp} = \frac{1}{N^\star_{pl}}\sum_j^{N^\star_{pl}} \cald{m}_{ij}\ ,
\end{equation}
and we can simply rewrite \autoref{eq:meanf} as a weighted average:
\begin{equation}
\cald{m}_{il} =  \sum_p \cald{m}_{ilp} w_{pl} \Biggm/ \sum_p w_{pl} \ ,
\label{eq:mcal}
\end{equation}
where the Poissonian optimal weights, $w_{pl} = N^\star_{pl}\sigma_p^{-2}$. Having assumed the RMS due to detector noise is exactly the same in each exposure, $i$ in the survey. 

Therefore, using \autoref{eq:meanf} and \autoref{eq:sigmap}, having summed over the $e_{ij}$, the tile miscalibrations, $e_{il} = \cald{m}_{il} - \true{m}_{l}$ are drawn from Gaussian, $N(0, \sigma_{l}^2)$ with:
\begin{equation}
\sigma_l^2 = \left(\sum_p w_{pl}\right)^{-1} \ . \\ \label{eq:sigma}
\end{equation}

Since the ubercalibration procedure uses only the differences between calibrated and true or mean values, we only ever generate the $e_{il}$ from a Gaussian with the width given in \autoref{eq:sigma}, having also generated the $w_{pl}$ using a Poissonian distribution of numbers of stars as described above.

Note that we have made several simplifying assumptions when we generated our synthetic calibrator averages. Before using this approach one should weigh its advantages against the disadvantage of losing the ability to calibrate the flat-field. In our case, this is in fact a desirable outcome as we have assumed the flat-field (or small-scale, pixel-to-pixel) self-calibration will be achieved otherwise.
We do not expect our relative analysis below to depend on these assumptions. 

\section{Self-calibration Procedure: Likelihood of Zero-points}\label{sec:likeli}

We would like to find the maximum of the posterior probability distribution of the calibration values, irrespective of the true fluxes of the calibrator stars. If $P(k_i, \true{m_j})~=~P(k_i)P(\true{m_j})$ is the $N_\exps \times N_\star$ multivariate Gaussian prior on our calibration and our stellar magnitude distribution, and $P(\mead{m_{ij}} | k_i, \true{m_j})$ the likelihood of the measured stellar magnitudes given the calibrations and the true values, we get the posterior probability for the set of the calibration and true values by Bayes to be:
\begin{equation}
P(k_i, \true{m_j} | \mead{m_{ij}})
= \frac{P(\mead{m_{ij}} | k_i,  \true{m_j}) P(k_i, \true{m_j})}{P(\mead{m_{ij}})} \ .\\
\end{equation}
However, we are only interested in the calibrations, $k_i$ of exposures $i$. 

Similarly to the \citet{Padmanabhan:2008} analysis, we would like to marginalise over the true values of the stellar magnitudes to obtain:
\begin{equation}
P(k_i | \mead{m_{ij}})
= \frac{P(\mead{m_{ij}} | k_i) P(k_i)}{P(\mead{m_{ij}})} \ .
\end{equation}
It can be shown that using $\cald{m_{ij}}~=~\mead{m_{ij}}-k_i $ and \autoref{eq:estimator} below gives the $-2\ln$ of the final posterior (modulo additive constants) as:
\begin{equation}
\chi_\eff^2(\{k_i\} | \mead{m_{ij}}) =
\sum_{ij} \frac{\left(\cald{m_{ij}}- \cald{\overline{m}_j}\right)^2}{\sigma_{ij}^2}
+ \sum^{N_\exps}_i\left(\frac{k_i}{\sigma^\textrm{k}}\right)^2  \ .
\label{eq:starposteri}
\end{equation}
where we have rewritten the posterior as a function of the average over $N_j^\exps$ exposures\footnote{Note that in order to compensate for the decreased scatter around the measured mean, the denominator should be multiplied with the Bessel factor $\sigma_{ij}^2 \rightarrow \left(\frac{N_j^\exps-1}{N_j^\exps}\right)\sigma_{ij}^2$.} of each star, $j$ in order to obtain an estimator of the true stellar magnitude, $\true{m_j}$:
\begin{equation}
\cald{\overline{m}}_j = \frac{1}{N_j^\exps} \sum_i^{N_j^\exps} \cald{m_{ij}} \ .\label{eq:estimator}
\end{equation}
If the calibration correctly accounts for the systematic bias in the measurement, this estimator will by definition be unbiased. This simplifies the minimisation routine as we do not need to explicitly estimate calibrator brightnesses, but instead automatically marginalises over their values. The intermediate steps can be followed in \autoref{app:posteri} in the Appendix.

Finding the minimum of this with respect to the set of $\{k_i\}$ gives us an unbiased estimate of the calibrations. However, the width of their distribution can be estimated using the second derivative of \autoref{eq:starposteri} only up to a Bessel correction factor.

To speed up the calculation of the $\chi^2_\eff$ in \autoref{eq:starposteri} by compressing our calibrator data to only retain the information relevant to the calibration we wish to perform, we use the quantities derived in \autoref{sec:patches}. As we do not need the measurements of individual stars, but rather derive the calibrations from the comparisons of measurements between the same stars in different exposures, we can perform an optimally weighted averaging over a contiguous fragment of the sky that is imaged by the same set of exposures. This rectangle-shaped overlap tile must by definition be smaller than a detector image. Note that this only works because we are interested in fixing the exposure-to-exposure variation in the calibration, leaving any pixel-to-pixel variation in the flat field to be constrained otherwise. We therefore split up the first sum in \autoref{eq:starposteri} into $N_\overlap$ sets of stars in the full survey:
\begin{equation}
\chi_\textrm{L}^2(\mead{m_{ij}} | \{k_i\}) = 
\sum_i^{N_\exps} \sum_l^{N_\overlap} \sum_j^{N_l^\star} 
	\frac{\left(\cald{m_{ij}} - \cald{\overline{m}_j}\right)^2}{\sigma_{ijl}^2}\ ,
\end{equation}
where $N_l^\star$ is the number of stars, $j$ in overlap $l$, $\cald{\overline{m}_j}$ is the mean of all calibrated measurements, $i$ of the stellar magnitude, and $\sigma_{ijl}~=~\sigma_{ij}$ when star $j$ appears in overlap $l$, but $\sigma_{ijl} = \infty$ otherwise.

Then we can calculate the likelihood contribution to the effective $\chi^2$ as:
\begin{equation}
\chi_\textrm{L}^2\left(\mead{m}_{il} | \{k_i\right\}) = 
\sum_l^{N_\overlap} \sum_i^{N_l^\text{exp}}
	\frac{\left(\cald{m_{il}} - \cald{\overline{m}_l}\right)^2}{\sigma_{l}^2}\ ,
\label{eq:likelihood}
\end{equation}
where $\cald{\overline{m}}_l = \sum_i \cald{m}_{il} / N_l^\text{exp}$ and where now $N_l^\text{exp}$ is the number of exposures that contain the tile $l$. 

It can be shown 
that minimising the $\chi^2$ in \autoref{eq:starposteri} with either of the $\chi^2_\textrm{L}$ will give the same set of calibrations, $k_i$ with a modified distribution in non-minimum solutions. Therefore, our final $\chi^2$ equation is:
\begin{equation}
\chi_\eff^2(\{k_i\} | \mead{m_{li}}) =
\sum_l^{N_\overlap} \sum_i^{N_l}
\frac{\left(\cald{m_{il}}- \cald{\overline{m}_l}\right)^2}{\sigma_l^2}
+ \sum^{N_\exps}_i\left(\frac{k_i}{\sigma^\textrm{k}}\right)^2  \ ,
\label{eq:overposteri}
\end{equation}
where in the first term the dependence on $k_i$ is in the $\cald{m_{il}}$.

Finding the set, $\{k_i\}$ that minimises this equation (using \autoref{eq:mags}) therefore gives an optimal set of calibrations to use to correct the magnitude zero-points of the survey, $\{i\}$.

\section{A Simple Model for the Spectroscopic Component of the Euclid Survey}\label{sec:Euclid}

The Euclid satellite mission will consist of 3 channels using the same 1.2 meter telescope: visual imaging, NIR imaging photometry and slit-less NIR spectroscopy. The latter two will be performed using an instrument with one set of NIR detectors and filter-wheels to apply different photometric filters and spectroscopic grisms. The photometry will be in Euclid Y, J and H bands and the spectroscopy will cover the wavelength range of 12,500 - 18,500 \AA. In this paper, we consider only the relative calibration of the spectroscopic channel, but this analysis could in principle easily be extended to the other two.

Each spectroscopic \textbf{exposure} will have a FOV covered by a $4\times4$ array of H2RG NIR detectors \citep{Beletic:2008}. The baseline survey strategy will be step-and-stare with slightly overlapping adjacent pointings, each pointing will be dithered 3 times to cover each field with approximately 4 dithered exposures.

To test calibration strategies, we construct a smaller, simplified model of the spectroscopic component of the Euclid survey. In this work we consider a realisation of the NISP instrument. The focal plane is made up of $N_\textrm{det}=16$ \textbf{detector} squares. Each \textbf{pointing} of the telescope includes four dithered exposures of the entire focal plane and therefore generates a set of four four-by-four arrays of squares. Each detector is made up of 2040-by-2040 pixels, which corresponds to each detector covering an area of 612-by-612 squared arc-seconds\footnote{Note that we round the detector size to the nearest $100\arcsec$ in our model for simplicity.} on the sky at the NISP magnification. The chip-gaps between detectors correspond to 50 and 100 arc-seconds horizontally and vertically respectively. We furthermore disregard pointing errors and assume that each new pointing of the telescope starts exactly a chip-gap away from the edge of the first exposure of the previous pointing. This is conservative in terms of calibration as it is likely that further overlaps will result from a denser tiling actually applied. Therefore, if the detector width, $x=y=612\arcsec$, and the chip-gaps, $x_\text{gap} = y_\text{gap}/2 = 50\arcsec$, then one full focal plane exposure covers approximately $A_\textrm{exp} = N_\textrm{det}(x+x_\text{gap})(x+y_\text{gap}) \approx 0.58$ deg$^2$ on the sky.

\subsection{Pointing}

\begin{figure}
\centering
\includegraphics[width=\linewidth]{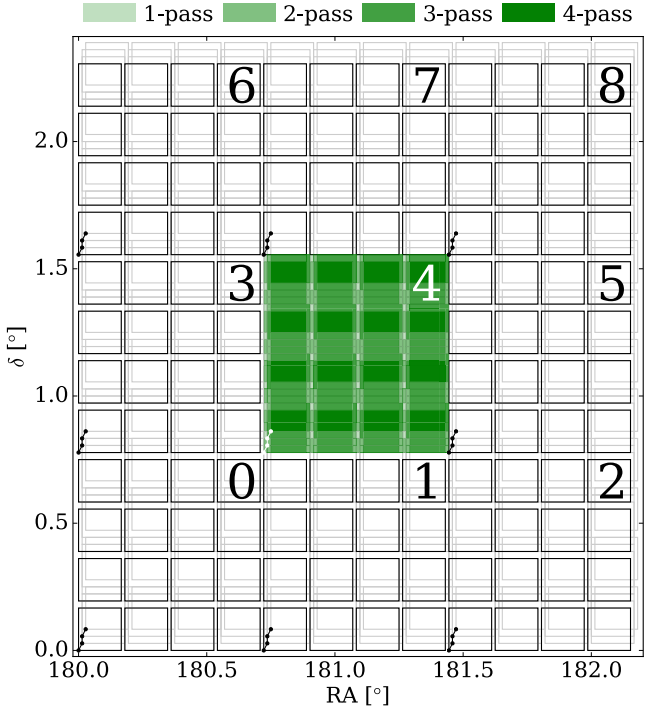}
\caption{9 adjacent pointings, each containing 4 dithers, under the unnatural (yet conservative) assumption that the sky is flat. Each pointing is labeled with a number 0 - 8 and consists of 16 detector imprints. In this plot, pointing \#4 is highlighted in green, showing the $n$-pass coverage. Each square corresponds to a single detector, the first exposure of each pointing is highlighted in black, the dithered exposures are shaded grey. The black (or white) connected dots indicate the first pixel of the dithers of each pointing.}
\label{fig:pattern}
\end{figure}

We consider a survey with $N_\textrm{point}$ pointings, each dithered $N_\textrm{dither}$ times. If $N_\textrm{dither}~=~1$ and $N_\textrm{point}~=~9$, the setup looks like the black pattern in \autoref{fig:pattern}. In this figure we emphasise only one dither for clarity, however in the rest of the paper we consider the default of $N_\textrm{dither} = 4$ \citep[matching ][]{Laureijs:2011}. Each pair of $(m,n)~\in~\{1...N_\textrm{point}\},~\{1...N_\textrm{dither}\}$ respectively corresponds to an exposure, $i$. 

As a simplification, we consider the sky to be flat, therefore the only source of overlaps is the dithering, having also assumed an infinite pointing precision. There is no overlap between adjacent pointings due to would-be Euclidean square tiling of the celestial sphere. Furthermore, we assume no overlaps due to pointing error, which do not have to be negligibly small in a space-based survey like Euclid, where they are expected to be as large as $10\arcsec$ at $3\sigma$. As is shown in \citet{Holmes:2012} and below, irregularities in the geometry of overlaps \textit{improve} the power of the ubercalibration procedure as this ensures that very different regions of the focal plane can be sampled by the same calibrators. Therefore, since our assumptions described here increase the regularity, we can be satisfied that we are making a slightly conservative estimation of the power of such self-calibration. 

\subsection{Dither Patterns and Coverage}
The exposures will be slightly dithered in order to cover the parts of the sky that would otherwise fall into the gaps between detectors. Therefore the coverage of the sky will produce a repeating pattern of 4-, 3-, 2-, 1-, and 0-pass regions.

\begin{table*}
    \begin{tabular}{|l||rr|rr|rr|r|r|r|}
        \hline
        pattern & $d_{x1}$ & $d_{y1}$ & $d_{x2}$ & $d_{y2}$ & $d_{x3}$ & $d_{y3}$ & $r$ & $\abs{\vec{d}}$ & $D$\\
        \hline
        S       & 50\arcsec & 100\arcsec &   0\arcsec & 100\arcsec &  50\arcsec & 100\arcsec & 1.0 & 316.23\arcsec & 323.61\arcsec\\
        J       & 50\arcsec & 100\arcsec &   0\arcsec & 100\arcsec &   0\arcsec & 100\arcsec & 1.0 & 304.14\arcsec & 311.80\arcsec\\
        N       & 50\arcsec & 100\arcsec &  50\arcsec &   0\arcsec &  50\arcsec & 100\arcsec & 1.0 & 250.00\arcsec & 273.61\arcsec\\
        R    & 50\arcsec & 100\arcsec &  50\arcsec &   0\arcsec &  50\arcsec &   0\arcsec & 1.0 & 180.28\arcsec & 211.80\arcsec\\
        O     & 50\arcsec &   0\arcsec &   0\arcsec & 100\arcsec & -50\arcsec &   0\arcsec & 1.0 & 100.00\arcsec & 200.00\arcsec\\
        X       &  0\arcsec &   0\arcsec & 50\arcsec & 100\arcsec &   0\arcsec &   0\arcsec & 1.0 & 111.80\arcsec & 111.80\arcsec\\
        \hline
    \end{tabular}
    \caption{Table of displacements in arc-seconds for each dither pattern considered in this paper with the default $d_x=50\arcsec$ and $d_y=100\arcsec$. The $d_{xn}$ and $d_{yn}$ are the components of the $(n+1)^\text{th}$ dither position vector in arc-seconds with respect to the pointing position. The fourth column, $r$ is the scale of the dither steps with respect to the reference model in \citet{Laureijs:2011}. The penultimate column, $\abs{\vec{d}}$ is the magnitude of the total displacement from the first to the final dither. The last column, $D$ is the total telescope travel from first to last dither.}
    \label{tab:patterns}
\end{table*}
We parametrise this dithering pattern with an array of three 2-dimensional dither vectors. Each 2-dimensional vector in \autoref{tab:patterns} describes the displacement from the previous dither position, starting at the position of the pointing:
\begin{equation}
d = \left[(d_{x1}, d_{y1}), (d_{x2}, d_{y2}), (d_{x3}, d_{y3})\right]\ .
\end{equation}
For example, what we call the  \textbf{S-pattern} for 4 dithers would look like the following 3 consecutive displacements from $(0.0, 0.0)$:
\begin{equation}
d(d_x,d_y) = \left[(d_x, d_y), (0.0, d_y), (d_x, d_y)\right]\ ,
\end{equation}
with $d_x = 50\arcsec$ and $d_y=100\arcsec$ in all our reference scenarios, which is chosen to exactly cover the inter-detector gaps in the focal plane.
Keeping the horizontal, $d_x$ and vertical, $d_y$ displacements constant in the array, i.e.
\begin{equation}
d_{xn} = d_{x1}\,\ \text{and}\ d_{yn} = d_{y1} \ ,\ \text{with}\ d_y = 2d_x \label{eq:dithersize}
\end{equation}
is something we do throughout this analysis to constrain the number of patterns we need to explore. 

However, in the sections below, we do vary the scaling, $r$ of the total dither vector. Then
\begin{equation}
\vec{d} = \vec{d}(d_x=r x_\text{gap},\ d_y=r y_\text{gap})\ ,
\end{equation}
and so $r$, or equivalently the x-shift of the dithers, $d_x$ fully constrains the size of a given pattern.

In addition, we calculate the total distance the telescope must move during dithering as
\begin{equation}
D = \sum_{n=1}^{N_\textrm{dither}-1} \sqrt{d_{xn}^2+d_{yn}^2}\ .
\end{equation}\\

\begin{figure*}
\centering
\includegraphics[width=\linewidth]{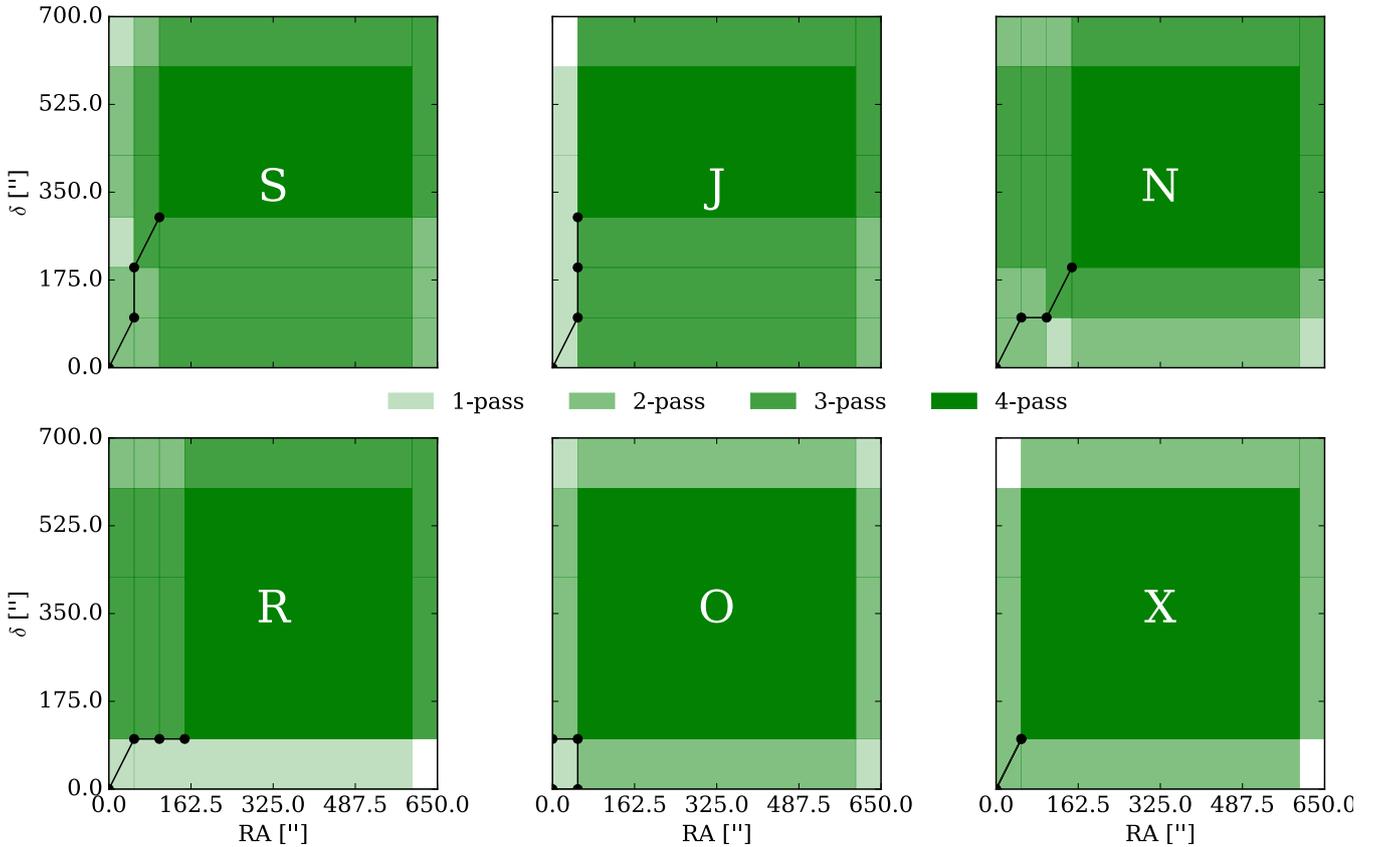}
\caption{The $n$-pass overlap footprints of the S, J, N, R, O, and X dithering patterns respectively from left to right and top to bottom. See also \autoref{tab:patterns}. What is left white is not observed.}
\label{fig:patterns}
\end{figure*}

In this paper we consider 6 different patterns named approximately according to the shape that a single pixel on the focal plane describes during the full dithering of one pointing. The basic patterns are S and J, the latter having been presented in \citet{Laureijs:2011} and used in our plots as a reference. It is worth recalling that the reference solution is only one among those which fulfill all the constraints and that was used to show the mission feasibility. In fact, this solution is likely to change after the corresponding optimisation phase studies are completed.

Mirroring S and J across the diagonal gives N and R respectively. To that we add two very simple control patterns, O and X. 
These 6 patterns could be split into 3 groups: S and N, J and R, O and X in order of decreasing complexity. We choose these patterns on the basis of how they connect survey area in the reference scenario. S and its horizontal version, N in the reference scenario do not leave any gaps in the coverage and connect not only vertically, but also horizontally to adjacent detector tiles and pointings. J and it's horizontal version R only connect vertically adjacent tiles. O and X are studied simply as potentially overly simple patterns. The dither vectors for these patterns can be seen in \autoref{tab:patterns}. Note that in the case of dither steps larger that those in the reference scenario, the survey quickly becomes entirely connected and the difference between the patterns diminishes.

We show the regions of 4, 3, 2, 1 and 0-pass coverage in \autoref{fig:patterns} for a smallest repeating fragment (of the size of a detector and one set of gaps) of the central region in a survey of 9 tiled pointings, each dithered 4 times in the default size ($r=1.0$) of each of the 6 patterns that we consider in this paper. 
We choose them such that at a scale of $r=1.0$ all the patterns have the same x-step, $d_x = 50\arcsec$ (when not set to 0 by the pattern definition) and require comparable amounts of spacecraft consumables, indicated by the total telescope movement required to make the full dithers, $D$. We wish to find the pattern and size that gives us the best returns in the calibration for a similar amount of resources and without negatively impacting the coverage.

Different dither patterns therefore result in different fractions of coverage (see \autoref{tab:coverage}). Note that there is a difference to the percentage reported in Table 4.8 of the Euclid Red Book \citep{Laureijs:2011}. This is likely explained by the different assumptions about the inter-pointing spacing in our simple model. Whereas the Red Book assumes no inter-pointing spacing, in this paper we repeat the inter-detector gaps when tiling the first exposures of different pointings. This can for example be seen in Figure 4.18 in the Red Book. Secondly, we do not consider any pointing errors in our model survey.

\begin{figure}
\centering
\includegraphics[width=\linewidth]{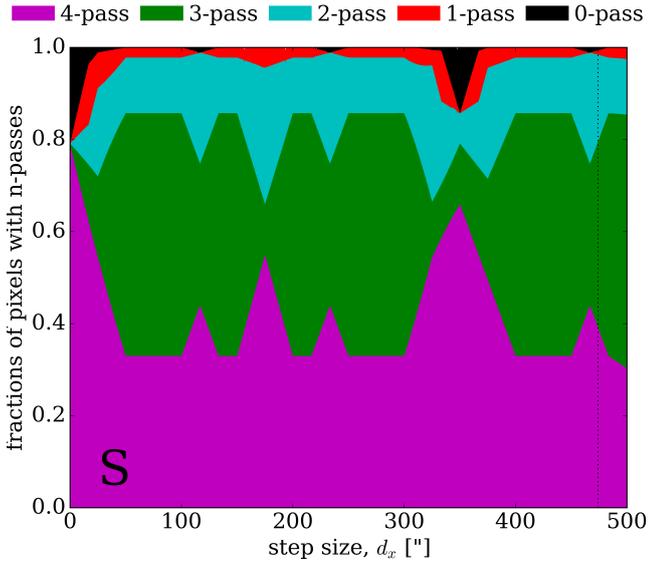}
\caption{Fractional coverage as a number of passes with dither vector scaling, $r$ for the S-pattern of dithering. An $r=1.0$ corresponds to the first dither displacement equal to the detector gap, i.e. $d_x=50\arcsec$. The magenta, green, cyan, red and black areas, stacked from bottom to top, correspond respectively to 4-, 3-, 2-, 1- and 0-pass coverage of the survey region. The vertical dotted black line marks where our survey geometry algorithm becomes less reliable, since the last dither moves beyond the first dither of the pointing.}
\label{fig:coverage}
\end{figure}

As also seen in \autoref{fig:pattern} we calculate this by constructing a survey mask of $3\times3$ pointings using the Mangle code \citep{Swanson:2008}. Using the same code we find the overlap regions that are covered with 4, 3, 2, 1, and 0 passes. For this calculation we discard all the overlap regions that lie outside of the first exposure of the central pointing, to which we count both the area of the 16 detectors as well as the gaps between them and half the gaps around the exposure. In our setup this pattern of passes repeats exactly, so even the region of one detector and its gaps (see also \autoref{fig:patterns}) suffices for this calculation. We add together the areas in square degrees covered by each number of passes. 
In \autoref{fig:coverage} we show how the coverage changes with the dither scale, parametrised by the x-shift of the dithers, $d_x$ for the S-pattern, $d_x=50\arcsec$ being the optimal choice to maximise 3- and 4-pass coverage with minimal total displacement.
\begin{table}
    \begin{tabularx}{\linewidth}{|X||r|r|r|r|r|}
        \hline
        pattern & 0-pass & 1-pass & 2-pass & 3-pass & 4-pass \\
        \hline
        J       &  1.06\%&  6.49\%&  3.18\%& 52.06\%& 37.20\%\\
        S       &  0.00\%&  2.12\%& 11.80\%& 52.19\%& 33.89\%\\
        R    &  1.06\%& 12.98\%&  3.18\%& 32.59\%& 50.19\%\\
        N       &  0.00\%&  2.12\%& 18.29\%& 39.21\%& 40.38\%\\
        O     &  0.00\%&  4.24\%& 34.71\%&  0.00\%& 61.05\%\\
        X       &  2.12\%&  0.00\%& 36.83\%&  0.00\%& 61.05\%\\
        \hline
    \end{tabularx}
    \caption{Table showing percent coverage of the survey region with 0, 1, 2, 3 and 4 overlapping exposures for different dither patterns with the default dither size, i.e. $d_x=50\arcsec$ and $d_y=100\arcsec$.}
    \label{tab:coverage}
\end{table}\\

Note that the parameters and pattern of \citet{Laureijs:2011} were chosen on the basis of requirements other than \textit{ubercalibration}. However this paper is only concerned with this method and therefore does not consider the effects of using different dither patterns beyond the changes in coverage and the effectiveness of relative calibration from overlaps.

\subsection{Calibrator Stars}\label{sec:stars}

\begin{figure}
\centering
\includegraphics[width=\linewidth]{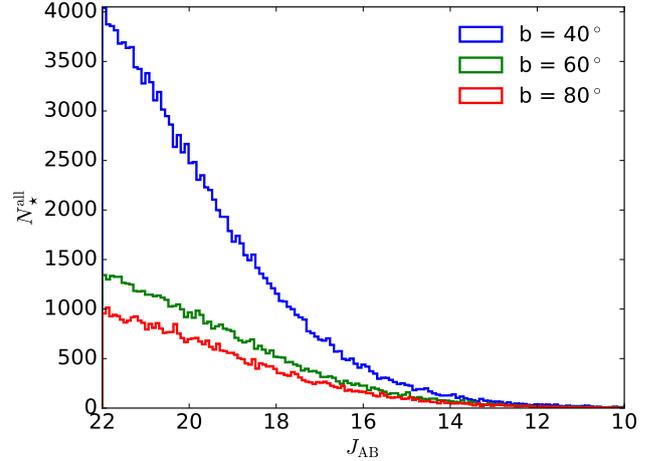}
\caption{Histogram of J magnitude (AB photometric system) of Trilegal stars in a 10 deg$^2$ patch of the Milky Way. All stellar types are shown, giving 235,689 stars in total at three different galactic latitudes, $b = 40^{\circ}, 60^{\circ}, 80^{\circ}$.}
\label{fig:stars}
\end{figure}

We assume relative calibrations are made using the brightness of stars in overlap regions. We conservatively assume that no extended objects can be used. We use the \href{http://stev.oapd.inaf.it/cgi-bin/trilegal}{Trilegal} \citep{Vanhollebeke:2009,Girardi:2005} model to generate the calibrator star J magnitudes at different galactic latitudes (see \autoref{fig:stars}) in the AB photometric system, \citep{Oke:1983}. For the analysis, we pessimistically choose to carry out our computations with a low density of bright stars, typical of the high galactic latitude sky at $b=80^\circ$. To obtain a robust estimate for this density, we have extracted a virtual 10 deg$^2$ field. The 10 deg$^2$ patch is generated from five 2-deg$^2$ patches at the same galactic latitude and random galactic longitude to calculate the average surface number density and signal-to-noise ratios (SNR). The results for stellar types F, K, and M are shown in \autoref{tab:Nstars} and \autoref{tab:snrs}.
\begin{table}
\begin{tabularx}{\linewidth}{|X|X||l|X|}
\hline
$J_\textrm{AB}$ limit & $n^\star$ & $J_\textrm{AB}$ bin centre & $n^\star_p$ \\
\hline\hline
$<$12.0 &   267 & 12.0 &   92 \\
$<$13.0 &   550 & 13.0 &  192 \\
$<$14.0 &  1028 & 14.0 &  319 \\
$<$15.0 &  1892 & 15.0 &  602 \\
$<$16.0 &  3402 & 16.0 &  975 \\
$<$17.0 &  5949 & 17.0 & 1563 \\
$<$18.0 &  9719 & 18.0 & 2343 \\
$<$19.0 & 15303 & 19.0 & 3321 \\
$<$20.0 & 23031 & 20.0 & 4288 \\
\hline
\end{tabularx}
\caption{Numbers of all the F, K and M-type stars with magnitudes in a 10 deg$^2$ patch at the galactic latitude of 80 degrees; cumulatively, left and binned, right.}
\label{tab:Nstars}
\end{table}

We check how realistic these models are with respect to the real sky as measured by the VIPERS survey \citep{Garilli:2014}. We compare the numbers of stars and the scatter in their brightness in the same magnitude bins as those in \autoref{tab:Nstars} and \autoref{tab:snrs}, this time generated for a galactic latitude of $b = -57^\circ$, with a 5-deg$^2$ patch of the 2-hour W1 VIPERS field. The star-galaxy separation of VIPERS is at magnitude $i_\textrm{AB} = 21.0$, which yields 11,369 stars in the VIPERS field. In the VIPERS control sample, the star-galaxy separation was carried out using purely the size of the image without other assumptions for stars brighter than $i_\text{AB}=21.0$, therefore it is extremely reliable for these relatively bright objects \citep{Guzzo:2014}. For stars fainter than $i_\text{AB}=21.0$, the loci of stars and galaxies start to merge because of the noise, and the separation becomes less reliable. Therefore we use only the stars brighter than this limit.
Comparing to the Trilegal model, which gives us 11,124 with the same magnitude cut, in the same field-size, we are satisfied that the numbers are close to reality, erring on the conservative side.

We then calculate the SNR from simulations of the NISP using the TIPS code \citep{Zoubian:2014} and the Euclid baseline parameters. 
We use the TIPS code to generate ``measured'' two-dimensional slit-less spectra from ``true'' one-dimensional stellar spectra using the \citet{Pickles:1998} spectral templates. 
We measure the scatter of the ``observed'' stellar fluxes around the ``true'' fluxes given by the model from the \citet{Pickles:1998} library. We do this by generating a number of simulations of observations of each star. 
Using the Imodel \citep{Imodel} pipeline we simulate the detection and extraction of the one-dimensional ``measured'' spectrum of each star, having subtracted the template continuum so as not to include real continuum variations in the calculation of the noise RMS. 
Stellar spectra were generated as to reproduce stars with a range of magnitudes, $J_\text{AB} = 14 - 20$. We generated 70 spectra for each one of the three spectral types, F, K, M and for each magnitude bin, making up a total of 1470 star spectra, which is the number of spectra that fit in a regular, well spaced grid avoiding overlaps.
Since the spectra are different for the three spectral types, the signal is different even if the stars have the same apparent $J_\text{AB}$ magnitude. Therefore we generated a range of spectral types in order to map out the variability in a real sky patch.

We then integrate the simulated single exposure spectra of selected stars over the spectral range of 13,000~-~14,000~\AA, over 102 pixels at wavelength sampling of 9.8 \AA/pixel\footnote{This resolution was specified in \citet{Laureijs:2011}, but note that this is \textbf{not} the expected final NISP wavelength resolution. This should not influence our relative results.}. The choice of integration wavelength range was intentionally minimal in order to be conservative to test that the method can still work even though we are only able to extract a partial calibrator spectrum, giving us a larger level of noise than if we had the full spectrum available.
As mentioned, we repeat this several times for each star. We measure the spread of the distribution of the measured integrated fluxes, $F_\textrm{int}$ for all the observations of each star, having kept the input ``true'' flux the same. This root-mean-square, $\textrm{RMS} = \Sigma_\textrm{int} / F_\textrm{int}$ is given in the \textit{RMS} column of \autoref{tab:snrs}. 
For comparison, in the \textit{expected SNR} column, we give the simple pixel-to-pixel scatter of the flux around the extracted continuum in the same wavelength range of single spectra. Since it does not include any error in background subtraction of flux calibration, the latter is therefore a little optimistic.

For a further comparison, to gauge the spread in the incoming signal related to the intrinsic spread of stellar fluxes, we consider the numbers of three stellar types, F, K and M \citep{Cannon:1925} at a fixed $J_\text{AB}$ magnitude. Their numbers are given on the right of \autoref{tab:Nstars}.

\begin{table}
\begin{tabularx}{\linewidth}{|c|c|X|X|l|}
\hline
$J_\textrm{AB}$ & type & RMS$_p$ & 1/RMS$_p$ & expected SNR$_p$ \\
\hline \hline
     & F & 0.00193 & 519.14 & 1335.70 \\
14.0 & K & 0.00161 & 620.77 & 1568.83 \\
     & M & 0.00108 & 928.66 & 1568.83 \\
\hline
     & F & 0.00252 & 397.27 &  834.26 \\
15.0 & K & 0.00218 & 459.62 &  931.34 \\
     & M & 0.00168 & 594.18 &  931.34 \\
\hline
     & F & 0.00356 & 281.01 &  442.22 \\
16.0 & K & 0.00348 & 287.26 &  487.80 \\
     & M & 0.00285 & 350.84 &  487.80 \\
\hline
     & F & 0.00642 & 155.85 &  209.71 \\
17.0 & K & 0.00616 & 162.23 &  217.57 \\
     & M & 0.00653 & 153.23 &  217.57 \\
\hline
     & F & 0.01594 &  62.74 &   80.55 \\
18.0 & K & 0.01513 &  66.11 &   87.80 \\
     & M & 0.01372 &  72.88 &   87.80 \\
\hline
     & F & 0.04102 &  24.38 &   32.86 \\
19.0 & K & 0.04386 &  22.80 &   35.70 \\
     & M & 0.03962 &  25.24 &   35.70 \\
\hline
     & F & 0.11044 &   9.05 &   13.53 \\
20.0 & K & 0.10742 &   9.31 &   14.58 \\
     & M & 0.07990 &  12.52 &   14.58 \\
\hline
\end{tabularx}
\caption{Signal-to-noise ratios of stellar fluxes using the Trilegal model as an input and simulating the measurement using the Euclid spectroscopic simulator, Imodel. The RMS is the root-mean-square deviation from the ``true'' stellar flux integrated over the 13,000~-~14,000~\AA\ range of the simulated observed flux in the same range. The expected SNR comes from the simple deviation of the pixel-to-pixel flux of individual spectra around their extracted continua. All these numbers are given for a high galactic latitude ($b=80^{\circ}$).}
\label{tab:snrs}
\end{table}

We also make the following optimistic assumptions. The stellar spectra are uncontaminated by adjacent objects on the sky that would normally be typical for slit-less spectroscopy. We do not include cosmic rays. We assume these effects would be minor and will degrade the stability of the calibration less than including more fainter stars will improve it. We make this assumption, because we choose mostly very bright stars that should be proportionally less contaminated by the fainter stars. Secondly, we assume that contamination by adjacent stars will be roughly constant exposure to exposure as the stellar density is roughly constant between adjacent pointings.

On average, this results in an uncertainty (not including the zero-point offsets) on the mean stellar magnitude in an overlap with an area, $A_l = 1\text{\ deg}^2$ of $\sigma_l(A_l) = 0.18$ mmag. This is the average value of what is used in the denominator of the first term in \autoref{eq:overposteri} to serve as the weight in the optimisation. It should be noted that this is not to be taken as the uncertainty on the final measured stellar magnitudes in the Euclid spectroscopic range. It is simply the expected purely statistical scatter between the measurements of the mean calibrator magnitude in an overlap patch of $1\ \text{deg}^2$.

\section{Simulation Results}\label{sec:results}

We generate calibrated measurements of the magnitudes of calibrator stars optimally averaged over each exposure, $i$ of each overlap tile, $l$ according to \autoref{sec:sims}. We use $\sigma^\text{k}_\text{f}$ as a metric of calibration improvement as it measures the final decreased scatter of the residual zero-points over the survey in magnitude.

We find the set of exposure zero-point calibrations ${k_i}$ that minimises \autoref{eq:likelihood} using a simple gradient descent method in the likelihood, updating our parameter set, $\{k_i\}$ until the $\chi^2_\text{eff}$ in \autoref{eq:overposteri} converges. The update rule uses the gradient of the likelihood multiplied by a learning rate, which simply equals to a quarter of the total error in an exposure, $i$, from all the overlap tiles, $l$ it covers:
\begin{equation}
k_i := k_i - \sum_l^{N^\text{o}_i}{\frac{\cald{m_{il}} - \cald{\overline{m}_{il}}}{\sigma_l^2}} 
  \bigg/ \sum_l^{N^\text{o}_i}{\frac{2}{\sigma_l^2}} \ ,
\end{equation} 
where $N^\text{o}_i$ is the number of useful overlap tiles covered by exposure $i$, and where now $\cald{\overline{m}_{il}}$ is the mean over all but the $i$-th exposure in overlap tile $l$.

We run two sets of simulations and calibrations. Both times we assume the large scale variations are initially completely unconstrained and the zero point of each subsequent exposure is drawn from a Gaussian with the width of $\sigma^\text{k}$ anew and is therefore independent of all other zero-points. 
We consider any other variation of the zero-point across one field-of-view to be due to the shape of the flat-field and calibrated otherwise. As mentioned throughout this paper, we also assume that there is no time variation of the shape of the flat-field, but we allow the average of the flat-field to vary freely.
We describe the results in \autoref{sec:exp} and \autoref{sec:det}.

\begin{figure}
\centering
\includegraphics[width=\linewidth]{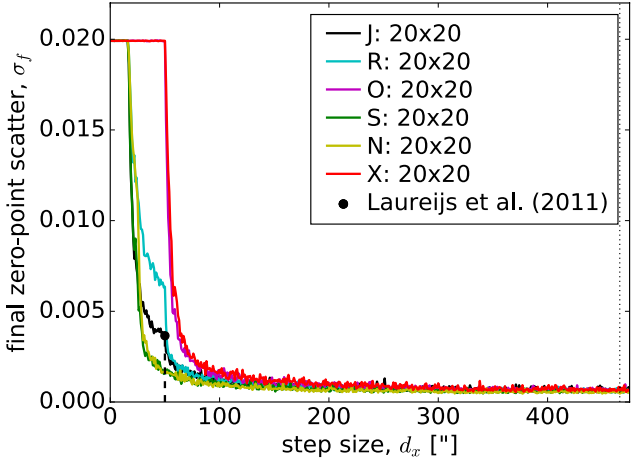} 
\caption{The final scatter of exposure zero-points for six different dither patterns described in this paper against the dither scale represented by the general horizontal displacement of the dither in the RA direction, $d_x$. The black, green, red, cyan, magenta and yellow line correspond to the so-called J-, S-, X-, R-, O-, and N-pattern  respectively. The black circle and dashed vertical line mark the reference strategy.}
\label{fig:expdx}
\end{figure}
\begin{figure}
\centering
\includegraphics[width=\linewidth]{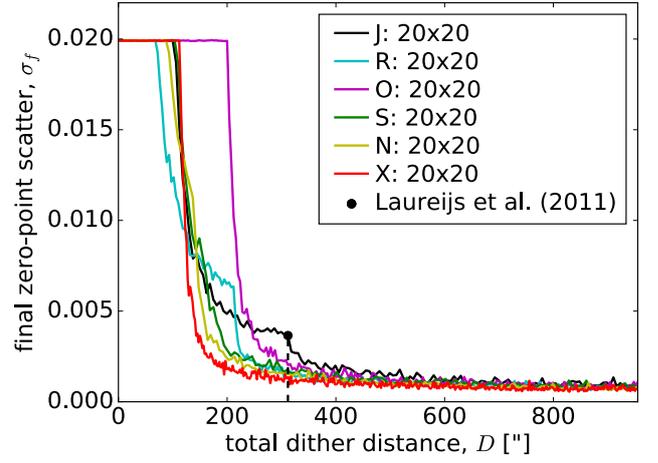}
\caption{The final scatter of exposure zero-points for six different dither patterns described in this paper against the total dither distance, $D$. The black, green, red, cyan, magenta and yellow line correspond to the so-called J-, S-, X-, R-, O-, and N-pattern respectively. The black circle and dashed vertical line mark the reference strategy.}
\label{fig:expr}
\end{figure}
\begin{figure}
\centering
\includegraphics[width=\linewidth]{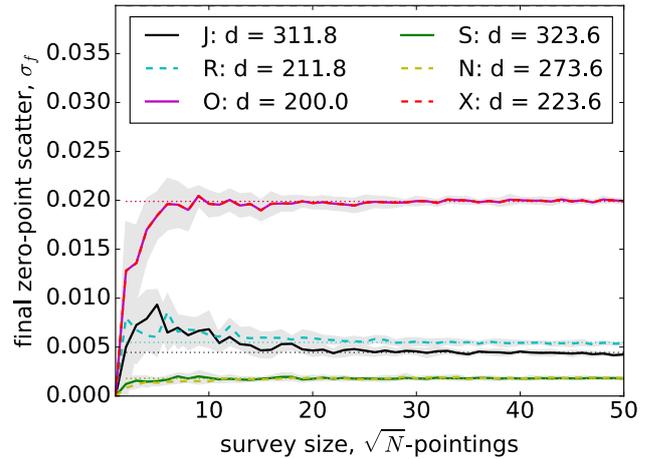}
\caption{The final scatter of exposure calibrations for four of the six patterns described in this paper for different survey sizes, but with the same dither scale, $r=1.0$, i.e. $d_x=50\arcsec$. We construct square surveys out of an integer number of pointings, given on the x-axis. We find 10 realisations of each configuration and plot their $1\sigma$ scatter (grey shaded region) and their mean (coloured lines). The dotted lines are the weighed averages across all the survey sizes with $N>1$. As before, the black, green, red, cyan, magenta and yellow line correspond to the so-called J-, S-, X-, R-, O-, and N-pattern respectively (some overlap).}
\label{fig:expsurvey}
\end{figure}

\subsection{Exposure-to-exposure}\label{sec:exp}

First we follow the procedure as described above and vary the calibrations \textbf{exposure-to-exposure}. We generate a single zero-point error for each full exposure of the focal plane, containing 16 detectors. This corresponds to our first choice of split between large- and small-scales. 
We take $\sigma^\text{k} = 40.0$~mmag (or 4\% in flux) between full focal plane exposures since this corresponds to a reasonable required minimum for this quantity. 

We collate the results of our exposure-to-exposure varying simulations in \autoref{fig:expdx}, \autoref{fig:expr}, and \autoref{fig:expsurvey}, where we plot $\sigma^\text{k}_\text{f}$, our quality metric that quantifies the uniformity of the calibrations across the survey after our procedure. We find that at the reference value of \citet{Laureijs:2011}, which corresponds to $d_x=50\arcsec$ in the J-pattern, the final scatter, $\sigma^\text{k}_\text{f} = 4.3$ mmag, meaning we are going from an exposure-to-exposure stability of $\sim4$\% to $\sim0.43$\% in flux after the relative re-calibration. However, at the same $d_x$, the S-pattern performs better and achieves a final scatter, $\sigma^\text{k}_\text{f} = 1.8$ mmag or $\sim0.18$\% in flux. We obtain these numbers by calculating an optimally weighted average over all the survey sizes considered (excluding the $N=1$ case). Because the S-pattern means a longer dithering vector with the same $d_x$, we consider an S-pattern with the same total dithering path as the nominal J-pattern, which has $d_x=48.17\arcsec$. This still provides a large improvement on the reference and results in a final flux calibration stability of $\sigma^\text{k}_\text{f} \sim 2.0$ mmag (see also \autoref{fig:expr}) in this very simple model.

To check that our results converge with the size of the survey and are statistically robust, we repeat the calculation by keeping the $d_x=50\arcsec$ constant, and changing the number of pointings in the square survey. In addition, we calculate several realisations of each survey size and calculate their scatter. It can be seen from \autoref{fig:expsurvey} that $\sigma^\text{k}_\text{f}$ converges for all the patterns as the survey size is increased.\\

\begin{figure}
\centering
\includegraphics[width=\linewidth]{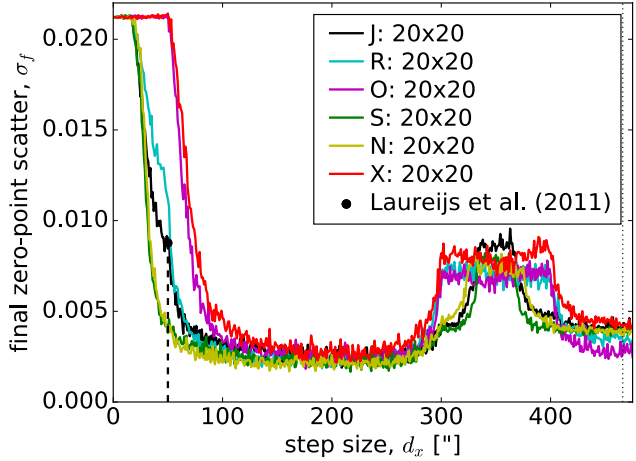}
\caption{The full exposure (averaged over the 16 detectors) calibration improvement for different dither scales allowing detector-to-detector variation of zero-points. The zero-points of each detector exposure are allowed to vary randomly.}
\label{fig:detdx}
\end{figure}
\begin{figure}
\centering
\includegraphics[width=\linewidth]{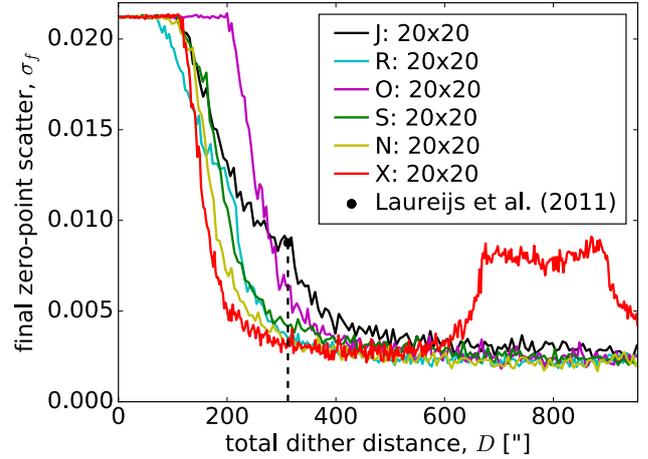} 
\caption{The full exposure (averaged over the 16 detectors) calibration improvement against the total dither distance, allowing each exposure with each detector to have its own randomly chosen zero-point.}
\label{fig:detr}
\end{figure}
\begin{figure}
\centering
\includegraphics[width=\linewidth]{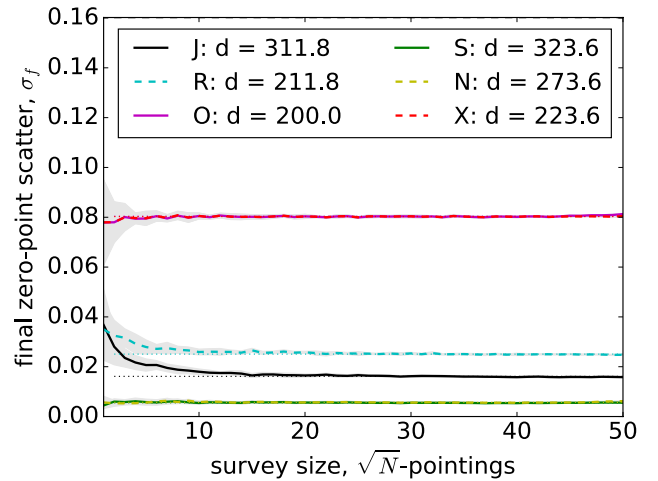}
\caption{The calibration improvement against the survey size, this time allowing each exposure with each detector its own randomly chosen zero-point. Note that the y-axis scale differs from \autoref{fig:expsurvey}, because we are plotting the detector-to-detector scatter. There are 16 detectors in each exposure in this set up.}
\label{fig:detsurvey}
\end{figure}

\subsection{Detector-to-detector}\label{sec:det}

We then consider the relative effectiveness of our dither patterns for the purposes of ubercalibration, if we make a different large- and small-scale split. Therefore, we run a second set of simulations, where we allow the calibration to vary \textbf{detector-to-detector}. The procedure is exactly analogous as in the first set, with the difference that now each exposure by each detector is treated as a new, independent exposure and indexed with $i$. 
In this second case, where we allow a scatter also between the 16 detectors in the focal plane, we appropriately adjust the detector-to-detector zero-point scatter to $\sigma^\text{k} = 160.0$~mmag (or 16\% in flux) to keep the initial total exposure-to-exposure scatter the same as in \autoref{sec:exp}.

This does not fundamentally change anything in our analysis. It is simply an alternative way to split up large and small scales. This step is rather pessimistic, since it assumes a complete ignorance about the relative zero points of all the detectors as well as the zero-point variation from dither to dither. We assume to only know the distribution of these zero-point errors, which we again assume to be Gaussian with a scatter of $\sigma^\text{k}$ as in \autoref{sec:exp}.

We plot the results of our simulations of the calibration procedure, where we have allowed the calibrations of the individual detector tiles to vary in \autoref{fig:detdx}, \autoref{fig:detr}, and \autoref{fig:detsurvey}. 
In \autoref{fig:detdx} and \autoref{fig:detr}, one can notice that the improvement of the survey uniformity at no dither displacement is not as good as a factor of $\sqrt{N_\text{dith}}=2$, from 40 mmag to 20 mmag as one may naively expect. This is most likely due to a combination of three factors. The first and probably dominant effect is that we are comparing the initial pure zero-point scatter, which is unknown, to the final, measured zero-point scatter, which in a way includes the measurement errors of the true mean stellar magnitudes. Secondly, we expect that the \textit{ubercalibration} procedure introduces correlations between the final zero-points whose exposures overlap with each other, which makes the calculation of expected zero-point scatter a little more complicated that a division with $\sqrt{N_\text{dith}}$. Finally, in these two plots, we are only looking at one realisation of the initial scatter of calibrators, which means that the exact value is sensitive to statistical fluctuations.
It is also apparent from \autoref{fig:detsurvey} that the calibration quality converges more quickly with survey size in this case than in \autoref{sec:exp}. This is likely to be simply due to having 16-times more data points. 

In this, more pessimistic scenario, the mean post-calibration zero-point scatter between full exposure zero-points (i.e. averaged over their 16 detectors) for the reference S-pattern is $\sigma^\text{k}_\text{f}~=~4.0$~mmag and $\sigma^\text{k}_\text{f}~=~7.7$~mmag for the J, both with $d_x=50\arcsec$. Taking the slightly smaller $d_x=48.17\arcsec$ in the S-pattern again gives $\sigma^\text{k}_\text{f}~\sim~4.8$~mmag.

In this case it is possible to see additional features around $d_{x1}~\approx~600\arcsec/2~=~350\arcsec$ i.e. half of width of detector $+~100\arcsec$ vertical gap, meaning that $d_{y1} = 700\arcsec$, which corresponds to a perfect vertical alignment (see \autoref{fig:patterns2} in the Appendix) of all dithers and therefore a large fraction of 0-pass coverage for all the patterns (for S, see \autoref{fig:coverage}). This feature indicates a decrease in uniformity (increase in $\sigma^\text{k}_\text{f}$), since the survey becomes less interconnected, which corresponds to a large 0-pass coverage, separating the different connected regions. See also \autoref{app:plots} in the Appendix for more plots. The additional feature in this scenario corresponds therefore to introducing a new typical scale of $\sim0.5$~detector + gap width.\\


\begin{table}
    \begin{tabularx}{\linewidth}{|l|X||r|r|r|r|r|r|}
        \hline
        pattern & r & J & S & R & N & O & X\\
        \hline\hline
        \multirow{2}{*}{exp-to-exp} & 1.0 & 4.3 & 1.8 & 5.4 & 1.9 & 19.9 & 19.9 \\ 
        						& 2.0 & 0.9 & 1.0 & 0.7 & 0.7 & 0.8 & 1.1 \\ 
	\hline
        \multirow{2}{*}{det-to-det} & 1.0 & 7.7 & 4.0 & 11.4 & 4.4 & 20.3 & 20.1\\ 
        						& 2.0 & 2.6 & 2.4 & 2.2 & 2.2 & 2.3 & 2.6\\ 
        \hline
    \end{tabularx}
    \caption{The scatter $\sigma^\text{k}_\text{f}$ in mmag between exposure zero-points at the reference scenario ($r=1.0$, $d_x=50\arcsec$, $d_y=100\arcsec$) in the upper rows, and the $r=2.0$ scenario in the lower rows, for all 6 types of patterns considered in this paper. The initial total exposure-to-exposure scatter, $\sigma^\text{k}=40$ mmag in both, the exposure-to-exposure (\ref{sec:exp}) and the detector-to-detector (\ref{sec:det}) case. The $r=1.0$ ($r=2.0$) are mean values calculated from 10 realisations of 50x50 (30x30) surveys for each scenario in the table.}
    \label{tab:results}
\end{table}
We summarise the above results at the reference scenario dither step ($dx = 50\arcsec$) in \autoref{tab:results}. We show the results from both \autoref{sec:exp} as ``exp-to-exp'' and \autoref{sec:det} as ``det-to-det'' for the 6 base dither patterns considered in this paper. In the ``det-to-det'' case we compute the scatter of the full exposure zero points, averaging over the zero-point of the 16 individual detectors in one exposure. This is the same procedure used to produce \autoref{fig:detdx} and \autoref{fig:detr}.

It is important to correctly split up the calibration into small and large scales. The present technique is good for intermediate-scale calibration, but something more extensive is required for the small scales. For example, one may calculate an average $\ln F$ for each pixel over all exposures to calibrate the pixel-to-pixel zero-points. This of course does not fix any time decay, where one should do the opposite and average over all the pixels in each exposure or detector and try to find long-term trends. Here we are studying the intermediate step, calibrating the noisy exposure-to-exposure zero-point variations.

Note also that for all dither patterns in both sets of simulations the \textit{ubercalibration} improves as the dither step is increased up to the size of the section with a constant zero-point (exposure in \autoref{sec:exp} or detector in \autoref{sec:det}). This is because this increases the overlap between adjacent independent measurements of the calibrator flux, eventually connecting the entire survey together. However as the step is increased, more telescope consumables are used up. Therefore an optimal trade-off should be found and the pattern with the best trade-off should be used. Therefore, this paper shows that S is the optimal pattern for ubercalibration giving the best stability in exchange for the smallest telescope movement required.

\section{Conclusion}\label{sec:conc}

We have studied how the potential for undertaking an \textit{ubercalibration} style retrospective relative self-calibration based on matching the brightness of objects observed in overlap regions can be taken into account in survey strategy design. In particular, we have tested the efficiency of six dither patterns in determining a highly stable intermediate-scale calibration of our simple model of the spectroscopic component of the Euclid survey. 
In order to avoid fitting for individual stellar brightness and significantly increase the speed of calculation, we introduced a new simplification to the calculation of \citet{Padmanabhan:2008}, averaging over all the calibrator stars in all the relevant stellar populations, and fitting to the mean. We have shown that the method is equivalent with the standard assumptions of Gaussian errors.

We have conducted this analysis for the case of our simplified simulations. In order to apply our methods directly to data, a more complex study should be undertaken to test some of our assumptions. The Gaussian distributions of calibrator magnitudes would have to be one of the issues considered as would the calibration on intra-detector scales. The latter has been studied by \citet{Holmes:2012}, which we briefly revisited in \autoref{app:FF} of the Appendix. The former will have to account for outlier calibrators, most importantly variable stars, for example by replacing the means with medians, which are more resistant to extreme values. 

We have tested 6 dithering patterns, using 6 different dithering vectors (see \autoref{fig:patterns}). The best performing in terms of coverage and calibration stability improvements are S and J as they cover $\sim 86\%$ and $89\%$ of the sky with $>2$ passes. 
Considering only exposure-to-exposure zero-point variations, the S and J patterns in their reference size with $d_x=50\arcsec$ improve calibration stability from a distribution with RMS of 40.0 mmag to 1.8 mmag and 4.5 mmag respectively. The 40.0 mmag baseline corresponds approximately to $\sim$4\% fractional RMS variations in flux, and is the required limit for the instrument and telescope stability. We have shown how the \textit{ubercalibration} procedure can be used to achieve further calibration stability.

In addition, as \citet{Holmes:2012} show, larger increases in overlap between exposures with different zero-points generally tend to result in a larger improvement in the power of the \textit{ubercalibration} procedure. Larger overlaps generally result from larger dither steps up to the size of the repeating section of the exposure pattern. However, large dither steps cost more in fuel and time, so there is a clear trade-off with survey efficiency.

We have made some assumptions in this paper, whose impact on the effectiveness of \textit{ubercalibration} can be studied in the future.\\
Firstly, we only considered scaling the same six patterns up and down, but not changing the relative sizes of the steps. We believe this is sufficient to study the behaviour of the different dithering patterns and stick to the simplicity of limiting the number of patterns studied. \\
An important caveat in this analysis is that we only considered a flat, Euclidean sky, which is not appropriate for large surveys. However, we showed that the solution converges as we increase the size of the survey and do most of our analysis on surveys of $20\times20$ pointings, which span $\sim14.7\times15.8$ degrees on the sky. Furthermore, \citet{Holmes:2012} have shown that irregular overlaps improve the power of the \textit{ubercalibration} procedure, so we are making a conservative assumption.\\
Our study has only used the calibrator distribution of one galactic latitude with a relatively low density. However, our relative conclusions  are only weakly dependent on the density of calibrators. A higher calibrator density is expected to improve the results slightly.\\
We neglect any uncertainty in the flat-field. Including it would generally add parameters that need to be constrained due to the parametrisation of the flat-field. We have done this in order to focus on zero-point calibrations on large scales, which are most relevant to measurements of cosmological large-scale structure statistics. We assume that the intra-detector flat field parameters as well as their evolution will be constrained well using other calibration and self calibration techniques. \\
Studying the time evolution of the exposure-to-exposure and detector-to-detector variations is out of the scope of this work. It is however expected that models will be available to describe this evolution and therefore the calibration method would be designed to fit parameters of this model, rather than the calibrations of individual exposures, assuming that there is no other information. 

It is important to keep in mind that all these assumptions make this a toy model and the resulting large calibration stability should not be taken at face value. One should consider mostly the relative result when comparing the survey strategies. Further studies taking into account better models of instrument stability and better error modeling are needed to predict the power of \textit{ubercalibration} for Euclid.

We have presented a study of the optimisation of the Euclid NIR survey geometry with the sole focus on the large-scale calibration. We have shown that relative calibrations can be significantly improved by choosing certain steps and patterns. 
Because, as seen most clearly in \autoref{fig:expdx}, the \textit{ubercalibration} procedure can improve the uncertainty on the zero-points of exposures, it is worth including the proposed geometries into the optimisation studies of the Euclid survey, which will balance this with a wide range of additional limitations and requirements. Relative NISP calibration is only one of many goals the survey strategy must take into account, while still fulfilling a large number of requirements, which were not taken into account in this study.

Although we have considered an idealised scenario in this paper, calculations are underway and will be reported separately \citep{Scaramella:2017,Maiorano:2017} that consider a more realistic distribution of observations, when geometrical effects and noise are included. The qualitative conclusions of this work agree with our simplified calculations, and thus we believe our conclusions on the relative coverage fractions for the different dither strategies considered here are robust.

We leave to future work the study of the calibration procedure of surveys covering different regions of the galaxy as well as other large-scale effects. Among those are effects of the long term stability of the instrument, effects from solar system and galactic noise etc. In addition, we leave for future work the study of the impact of survey uniformity improvements due to different dither patters \citep[as in][ for the case of LSST]{Awan:2016} on clustering statistics.

We make our simulation code public on GitHub at \href{http://github.com/didamarkovic/ubercal}{\texttt{github.com/didamarkovic/ubercal}}.

\section*{Acknowledgements}

DM and WJP acknowledge support from the UK Science \& Technology Facilities Council through grants ST/M001709/1 and ST/N000668/1. WJP is also grateful for support from the European Research Council through grant 614030 Darksurvey, and support from the UK Space Agency through grant ST/N00180X/1. BG, LG, RS and MS acknowledge financial support by the Italian Space Agency, through contract ASI/INAF n. I/023/12/1. We thank Adriano Derosa, Carlo Burigana, Nikhil Padmanabhan, Julian Zoubian, Vincent Venin, and Florent Leclerc for useful discussions. In addition, we thank Rene Laureijs, Roland Vavrek and the Euclid Consortium. We acknowledge the use of the Astropy package for Python \citep{Astropy:2013}. Numerical computations were done on the Sciama High Performance Compute (HPC) cluster which is supported by 
the ICG, SEPNet and the University of Portsmouth.

\input{calibration.bbl}

\appendix

\newpage\mbox{}\newpage
\section{Tables of Symbols}

\setlength{\tabcolsep}{2pt}
\renewcommand{\arraystretch}{1.01}
\begin{table}
    \begin{tabularx}{\linewidth}{| l X r |}
        \hline
        \textbf{symbol} 				& \textbf{description} 											& \textbf{unit} \\
        \hline\hline
		$\alpha$				& equatorial coordinates right ascension								& $^\circ$ \\
		$\delta$				& equatorial coordinates declination									& $^\circ$ \\
		$b$					& galactic latitude												& $^\circ$ \\
		$x$					& horizontal coordinate on the detector								& $\arcsec$ \\
		$y$					& vertical coordinate on the detector									& $\arcsec$ \\
	\hline
	        	$x$					& width of the detector (clear from context)							& $\arcsec$ \\
		$y$					& height of the detector (clear from context)							& $\arcsec$ \\
		$x_\text{gap}$			& horizontal gap between detectors									& $\arcsec$ \\
		$y_\text{gap}$			& vertical gap between detectors									& $\arcsec$ \\
		$N_\textrm{point}$		& total number of fields in the survey									& \textit{none} \\
		$N_\textrm{dither}$		& number of times each field is dithered								& \textit{none} \\
		$N_\textrm{det}$		& number of detectors in one focal plane								& \textit{none} \\
		$m$					& indexes fields, max is $N_\textrm{point}$							& \textit{none} \\
		$n$					& indexes dithers, max is $N_\textrm{dither}$							& \textit{none} \\
	\hline
		$d$					& array of all dither displacements									& $\arcsec$ \\
		$d_{xn}$				& n$^\text{th}$ horizontal dither displacement							& $\arcsec$ \\
		$d_{yn}$				& n$^\text{th}$ vertical dither displacement							& $\arcsec$ \\
		$d_x$				& generic horizontal dither displacement								& $\arcsec$ \\
		$d_y$				& generic vertical dither displacement								& $\arcsec$ \\
		$r$					& scaling of the dither vector, i.e. of $d_x$ and $d_y$ from 50$\arcsec$ and 100$\arcsec$	& \textit{none} \\
		$\abs{\vec{d}}$			& total telescope displacement after dithers							& $\arcsec$ \\
		$D$					& total telescope movement during dithers								& $\arcsec$ \\
	\hline
		$J_\text{AB}$			& J magnitude (AB photometric system)								& mag \\
		$i_\text{AB}$			& i magnitude (AB photometric system)							& mag \\
		$\Sigma_\textrm{int}$	& integrated noise in the observed spectrum							& ADU \\
		$F_\textrm{int}$		& integrated signal in the observed spectrum							& ADU \\
		$p$					& indexes calibrator populations									& \textit{none} \\ 
		RMS$_p$				& root-mean-squared of the fluxes of $p$-type calibrators					& \textit{none} \\
		$\true{\sigma_p}$		& intrinsic scatter of magnitudes in $p$-type stars						& mmag \\
      	\hline
		$n^\star(\alpha, \delta)$	& surface number density of calibrator stars at a position $(\alpha,\delta)$ on the sky & deg$^{-2}$\\
		$n^\star(b)$			& average surface number density of calibrator stars at galactic latitude, $b$		& deg$^{-2}$ \\
		$n^\star_p$			& surface number density of $p$-type calibrator stars						& deg$^{-2}$  \\
      	\hline
		$i$					& indexes exposures, max is $N_\exps$								& \textit{none} \\ 
		$j$					& indexes calibrator stars, max is $N_\star$							& \textit{none} \\
		$l$					& indexes overlap tiles											& \textit{none} \\
	\hline
		$N_\star$				& total number of calibrator stars 									& \textit{none} \\
		$N^\star_l$			& number of all calibrator stars in the $l^\text{th}$ overlap tile				& \textit{none} \\
		$N_{pl}^\star$			& number of $p$-type calibrator stars in tile $l$									& \textit{none} \\
		$N_\exps$			& total number of exposures in survey								& \textit{none} \\
		$N_j^\exps$			& total number of observations of calibrator $j$							& \textit{none} \\
		$N^\textrm{pix}_j$		& area of the calibrator $j$ image									& pix \\
		$N_l^\text{exp}$		& number of exposures that contain the tile $l$							& \textit{none} \\
		$N_i^\text{o}$			& number of overlap tiles in exposure $i$								& \textit{none} \\
		$A_l$				& area of overlap tile $l$											& deg$^2$ \\
	\hline
    \end{tabularx}
    \caption{Table of sky, survey and dithering nomenclature used throughout this paper.}
    \label{tab:surpars}
\end{table}
\begin{table}
    \begin{tabularx}{\linewidth}{| l X r |}
        \hline
        \textbf{symbol} 				& \textbf{description} 											& \textbf{unit} \\
        \hline\hline
		$\true{\flux}_j$			& true flux	of calibrator $j$											& erg/s/cm$^2$ \\
		$\meadf{\flux_{ij}}$		& flux of calibrator $j$ measured in exposure $i$						& ADU \\
		$\cald{\flux_{ij}}$		& calibrated flux of calibrator $j$ in exposure $i$						& erg/s/cm$^2$ \\
		$\flux^0_i$			& mean dark current of exposure $i$									& ADU \\
		$K_i$				& multiplicative flux calibration mean of exposure $i$ to be determined		& \textit{none} \\
		$K_i(x,y)$				& multiplicative flux calibration at position $(x,y)$ in exposure $i$			& \textit{none} \\
		$G_i(x,y)$				& difference from $K_i$ at position $(x,y)$ in exposure $i$					& \textit{none} \\ 
	\hline
		$\true{m}_j$			& true magnitude of calibrator $j$									& mag \\
		$\true{m}_l$			& true mean magnitude of all calibrator in tile $l$						& mag \\
		$\mead{m}_{ij}$		& magnitude of calibrator $j$ measured in exposure $i$					& mag \\
		$\cald{\overline{m}_j}$	& mean of all calibrated magnitude measurements of calibrator $j$			& mag \\
		$\cald{m_{ij}}$			& magnitude of calibrator $j$ calibrated in exposure $i$					& mag \\
		$\cald{m}_{ilp}$			& magnitude of calibrator $j$ from population $p$, calibrated in exposure $i$	& mag \\
		$\cald{m}_{il}$			& mean calibrated magnitude of all stars in tile $l$ of exposure $i$			& mag \\
		$\cald{\overline{m}_l}$	& mean calibrated magnitude of all exposures in tile $l$					& mag \\
		$\cald{\overline{m}_{il}}$	& mean calibrated magnitude of all but the $i$-th exposure in tile $l$			& mag \\
		$e_{ij}$				& miscalibration of the magnitude of star $j$ in exposure $i$				& mag \\
		$e_{il}$				& miscalibration of the $l$-tile mean in exposure $i$						& mag \\
		$k_i$				& additive magnitude calibration of exposure $i$ to be determined			& mag \\
	\hline
		$\Sigma^\textrm{K}_j$	& scatter of calibrated flux measurements of calibrator star $j$				& erg/s/cm$^2$ \\
		$\sigma_{ij}$			& error on the measurement of magnitude $j$ in exposure $i$				& mmag \\
		$\sigma_{ijl}$			& error on the measurement of magnitude $j$ in overlap $l$ of exposure $i$
& mmag \\
		$\sigma_l$			& error on the measured mean magnitude $l$-tile						& mmag \\
 		$\sigma_i^{\textrm{e}^-}(x, y)$	& error in pixel at position $(x,y)$ in exposure $i$					& mmag \\
		$\sigma_i^{\textrm{e}^-}$	& error in a pixel in exposure $i$									& mmag \\
		$\sigma_p$			& scatter in measured magnitudes of $p$-type stars						& mmag \\
		$w_{pl}$				& Poissonian optimal weight of $p$-type calibrators in tile $l$				& mmag$^{-2}$ \\
       		$\sigma^\textrm{k}$		&  uncalibrated zero-point scatter 									& mmag \\
        		$\sigma^\textrm{k}_\textrm{f}$	& residual zero-point scatter 									& mmag \\
	\hline
		$P$					& probability (either posterior, prior or likelihood)						& \textit{none} \\
		$\chi^2_\text{L}$		& exponent of the likelihood										& \textit{none} \\
		$\chi^2_\text{eff}$		& exponent of the posterior										& \textit{none} \\
	\hline
    \end{tabularx}
    \caption{Table of calibration nomenclature used throughout this paper.}
    \label{tab:calpars}
\end{table}
\setlength{\tabcolsep}{6pt}
\renewcommand{\arraystretch}{1}
We summarise our notation in this section. To facilitate quick look-up we provide tables with brief descriptions and units. In \autoref{tab:surpars} we gather the symbols that parametrise the survey and dithering strategies. In \autoref{tab:calpars} we provide the parameters connected to the statistical description of the calibrators and calibrations.

\newpage\mbox{}\newpage

\section{Distribution of Miscalibrations}\label{app:dist}

We would like to show the relationship between the miscalibration, $e_{ij}$ defined in \autoref{eq:maximum} and it's flux equivalent $E_{ij}$, being the fractional flux miscalibration, which we define here, by writing the true flux of star $j$ as
\begin{equation}
\true{\flux_j} = \cald{\flux_{ij}} - \true{F_j}E_{ij}(\Sigma_{ij}/\true{F_j})\ .
\end{equation}
Therefore by \autoref{eq:flux}, the measurement of the true flux of star $j$ in exposure $i$ will be:
\begin{equation}
\meadf{\flux_{ij}} = \frac{1}{K_i}\true{\flux_{ij}}\left(1  + E_{ij}\left(\Sigma_{ij}/\true{F_j}\right)\right) + \flux^0_i \ ,
\label{eq:fluxnoise}
\end{equation}
with $\Sigma_{ij}/\true{F_j}$ being the standard deviation of the distribution of the fractional random relative miscalibration variable, $E_{ij}$. In the case of perfect ubercalibration, we would achieve $E_{ij}\rightarrow 0$ and $\sigma^E_{ij} \rightarrow 0$.

As a consequence of these assumptions, any difference between the flux measurements of two stars in the same exposure will be due to a random (close to Poissonian) noise from the detector in addition to intrinsic variations in the true stellar fluxes. We can characterise the detector-noise with the scatter $\Sigma_i^\ADU$ and the scatter in stellar brightnesses as $\Sigma_j^\star$. The better we understand these noise models and the behaviours of the corresponding quantities, the better the end result of this procedure. 
With this in mind, we later divide the variations of the second kind into those due to the different stellar types and a Gaussian distribution of magnitudes (not fluxes) among the stars of each type. We discuss this division above in \autoref{sec:sims}. 
The $\Sigma_{ij}$ is the expected scatter of measurements due to intrinsic flux scatter, the noise from the observation of each star (telescope and detector effects) and the error resulting from the extraction and analysis procedure. 
Since this quantity does not vary from star to star, but only the stellar type and exposure, we will say that if star $j$ belongs to population $p$, $\Sigma_j^\star=\Sigma^\star_p$, where the capital letters denote the relation of the quantities to flux rather than magnitude.

The first moment of the distribution of this variable can be derived from \autoref{eq:fluxnoise} by writing the true flux as:
\begin{eqnarray}
\true{\flux_j} 
&=& K_i^\textrm{t}\meadf{\flux_{ij}}\left[1+E_{ij}\left(\Sigma_{ij}/\true{\flux_j}\right)\right]^{-1} \nonumber\ , \\
\Rightarrow \true{m_j} &=& k_i + \mead{m}_{ij} + e_{ij}  \nonumber\ , \\
\textrm{where}\ e_{ij} 
&= &\true{m_j} - \cald{m_{ij}} \nonumber\\
&=& 2.5\log_{10}\left[1+E_{ij}\left(\Sigma_{ij}/\true{\flux_j}\right)\right] \ .
\end{eqnarray}
We can therefore find the width of the $e_{ij}$ distribution as:
\begin{equation}
\sigma_{ij} 
= \sigma^E_{ij} \left|\frac{de_{ij}}{dE_{ij}}\right|
= \left[\frac{\Sigma^\star_j}{\flux_j}\right] \frac{2.5}{\ln(10)\left(1+E_{ij}\right)} \ .
\end{equation}
We then find that averaging over all exposures, $i$ gives:
\begin{equation}
\left<\sigma_{ij}\right>_j \approx \frac{\Sigma^\star_j}{F_j} = \left[\left.\frac{\textrm{signal}}{\textrm{noise}}\right|_j\right]^{-1} \ ,
\end{equation}
if $E_{ij} << 1$ and $\left<E_{ij}\right>_j \approx 0$.

\section{Full Derivation of the Marginalised Posterior of the Multiplicative Exposure Flux Calibrations}\label{app:posteri}

We would like to calculate the posterior distribution in \autoref{eq:starposteri}, of the set of exposure calibrations $k_i$, given a set of measurements of calibrator magnitudes, $\mead{m_{ij}}$, marginalised over the true calibrator magnitudes, $\true{m_{j}}$.
This means we must first find the marginalised likelihood:
\begin{eqnarray}
P(\mead{m_{ij}} | k_i) &=& \int_{-\infty}^{\infty} P(\mead{m_{ij}} | k_i,  \true{m_j})P(\true{m_j})\ d\true{m}_j  \nonumber\\
&=& \text{const}\times\int_{-\infty}^{\infty} I(\true{m_j}) d\true{m}_j\ .
\label{eq:marged}
\end{eqnarray}
We assume the likelihood of measured stellar fluxes is a simple multivariate Gaussian and that the measured fluxes are independent, therefore the probability is simply a product of the individual probabilities, and taking the natural logarithm gives:
\begin{eqnarray}
\ln\left[P(\mead{m_{ij}} | k_i,  \true{m_j})\right]  =
&-& N_\exps\sum^{N_\star}_j\sum^{N_\exps}_i\ln(\sigma_{ij}\sqrt{2\pi}) + \ \ \ \ \ \ \ \\
&-& \frac{1}{2}\sum^{N_\star}_j\sum^{N_\exps}_i\frac{(\mead{m_{ij}} - (\true{m_j}+k_i))^2}{\sigma_{ij}^2} \nonumber\ , 
\end{eqnarray}
where the $\sigma_{ij}$ is also assumed to be known.
We can construct a prior on the true magnitudes expecting that they are distributed as a Gaussian with a mean of $\mu$ and a width $\tau_j$\footnote{The exact form of tho prior is not crucial for the rest of this procedure.}.
\begin{equation}
\ln\left[P(\true{m_j})\right] = -N_\star\ln(\tau\sqrt{2\pi}) 
- \frac{1}{2}\sum^{N_\star}_j\left(\frac{\true{m_j} - \mu}{\tau^2}\right)^2 
\label{eq:starprior}\ .
\end{equation}
From now on we drop all the constants and keep only terms that will be relevant in finding the maximum of the posterior. We are effectively working with the $\chi^2$ estimator as in \citet{Padmanabhan:2008}.
Then we can consider the $-2\ln$ of the integrand in \autoref{eq:marged}:
\begin{eqnarray}
-2\ln I(\true{m_j})
&=& 
\sum^{N_\star}_j\sum^{N_\exps}_i\frac{\left((\mead{m_{ij}} - k_i) - \true{m_j}\right)^2}{\sigma_{ij}^2}
+ \sum^{N_\star}_j\frac{(\true{m_j})^2 - 2\true{m_j}\mu}{\tau^2} 
\nonumber\\
&=& 
\sum_{ij}\frac{\left(\mead{m_{ij}}-k_i\right)^2}{\sigma_{ij}^2}
+ \sum_j M_j(\true{m_j}) \ ,
\end{eqnarray}
with
\begin{eqnarray}
M_j(\true{m_j}) &=& 
\left(\true{m_j}\right)^2
	\overbrace{
	\left[\sum_i\frac{1}{\sigma_{ij}^2}+\frac{1}{\tau^2}\right]
	}^{M_j^A}
- 2\true{m_j}
	\overbrace{
	\left[\sum_i\frac{\mead{m_{ij}} - k_i}{\sigma_{ij}^2} + \frac{\mu}{\tau^2}\right]
	}^{M_j^B}
 \nonumber\\
&=& 
\left(\true{m_j} \sqrt{M_j^A} - \frac{M_j^B}{\sqrt{M_j^A}}\right)^2
- \frac{(M_j^B)^2}{M_j^A} \ ,
\end{eqnarray}
having ``completed the square'' \citep{Bridle:2002}. Therefore, in order to marginalise over the true magnitudes, we must integrate over:
\begin{eqnarray}
\ln I(\true{m_j}) = &-& \frac{1}{2} \left[
	\sum_{ij}\frac{\left(\mead{m_{ij}}-k_i\right)^2}{\sigma_{ij}^2}
	- \sum_j \frac{(M_j^B)^2}{M_j^A}
	\right] + \nonumber\\
&-& \frac{1}{2} \sum_j \left(\true{m_j} \sqrt{M_j^A} - \frac{M_j^B}{\sqrt{M_j^A}}\right)^2 \ .
\end{eqnarray}
The second term in this equation is the only term that depends on $\true{m_j}$, therefore the other terms can come out of the integral as multiplicative constants. Then we will get an ordinary Gaussian interval, whose value will be independent of $M_j^B$ and therefore of $k_i$, which means that in the search for the maximum of the posterior with respect to $k_i$ it can be treated as a constant, i.e. ignored:
\begin{eqnarray}
P(\mead{m_{ij}} | k_i) &=&
\int_{-\infty}^{\infty} I(\true{m_j}) d\true{m_j} \\
&\propto& \exp\left\{-\frac{1}{2} \left[
	\sum_{ij}\frac{\left(\mead{m_{ij}}-k_i\right)^2}{\sigma_{ij}^2}
	- \sum_j \frac{(M_j^B)^2}{M_j^A}
	\right]
	\right\}
\ . \nonumber
\end{eqnarray}
Now that we have this general expression, we can loosen the prior on $\true{m_j}$:
\begin{equation}
\tau\rightarrow\infty
\ \textrm{ and therefore }\
\frac{(M_j^B)^2}{M_j^A}  
\rightarrow 
\left(\sum_i\frac{\mead{m_{ij}} - k_i}{\sigma_{ij}^2}\right)^2 \left[\sum_i\frac{1}{\sigma_{ij}^2}\right]^{-1}
\ . \nonumber
\end{equation}
Using that 
\begin{equation}
\sum_i \frac{\mead{m_{ij}} - k_i}{\sigma_{ij}^2} = \sum_i \frac{\overline{\mead{m_{ij}}-k_i}}{\sigma_{ij}^2}
\ ,
\end{equation}
we can write the marginalised likelihood:
\begin{eqnarray}
-2\ln P(\mead{m_{ij}} | k_i) &\propto&
	\sum_{ij}\frac{\left(\mead{m_{ij}}-k_i\right)^2}{\sigma_{ij}^2} + \\
	&-& \sum_j \left(\sum_i\frac{\mead{m_{ij}} - k_i}{\sigma_{ij}^2}\right)^2 \left[\sum_i\frac{1}{\sigma_{ij}^2}\right]^{-1}
\ . \nonumber
\end{eqnarray}
simply as an effective $\chi^2$:
\begin{equation}
\chi_\textrm{L}^2(\mead{m_{ij}} | k_i) = 
\sum_{ij} \frac{\left(\mead{m_{ij}}-k_i - \overline{\mead{m_{ij}} - k_i}_j\right)^2}{\sigma_{ij}^2}
\label{eq:starlikeli}
\end{equation}
We know that the absolute calibrations should be done to a certain quality, $\sigma^\textrm{k}$ and therefore we can also construct a prior to reflect this:
\begin{equation}
\ln\left[P(k_i)\right] =
-N_\exps\ln(\sigma^k\sqrt{2\pi}) 
- \frac{1}{2}\sum_i^{N_\exps}\left(\frac{k_i}{\sigma^k}\right)^2 
\ , \nonumber
\end{equation}
since by definition of \textit{relative} calibrations, taking the ensemble average over the full survey gives $\left<k_i\right> = 0$.
 And so, the $-2\ln$ of the final posterior (modulo additive constants) becomes:
 \begin{equation}
\chi_\textrm{eff}^2(k_i | \mead{m_{ij}}) =
\sum_{ij} \frac{\left(\mead{m_{ij}}-k_i - \left.\overline{\mead{m_{ij}} - k_i}\right|_j\right)^2}{\sigma_{ij}^2}
+ \sum^{N_\exps}_i\left(\frac{k_i}{\sigma^k}\right)^2  \ . \nonumber
\end{equation}


\section{Understanding the Features}\label{app:plots}

We wish to plot some more dither configurations at dither sizes that display clear features in \autoref{fig:expdx} and in \autoref{fig:detdx}, as well as coverage evolution with dither size for two of the the remaining five patterns, namely J and O. We choose to add the coverage of the J-pattern of \citet{Laureijs:2011} and we choose the O-pattern for its intuitive simplicity.
%
%
\begin{figure}
\centering
\includegraphics[width=\linewidth]{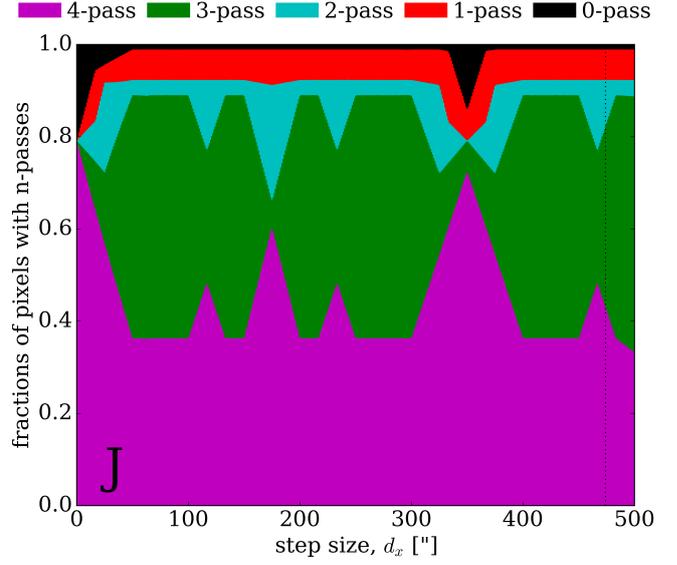}
\caption{Coverage fraction against dither size for the J-pattern. The same for the S-pattern can be see in \autoref{fig:coverage}.}
\label{fig:covsJ}
\end{figure}
%
%
\begin{figure}
\centering
\includegraphics[width=\linewidth]{\plotpath{plotcoverage-O}}
\caption{Coverage fraction against dither size for the O-pattern. The same for the S-pattern can be see in \autoref{fig:coverage}.}
\label{fig:covsO}
\end{figure}
%

\begin{figure*}
\centering
\includegraphics[width=0.9\linewidth]{\plotpath{plotsetup-patterns-350}}
\caption{The $n$-pass overlap tiles of the S, J, N, R, O, and X patterns from left to right and top to bottom, but with a dither scaling of $r=7.0$, corresponding to the dither step where the first dither aligns perfectly with the vertically adjacent row of detectors.}
\label{fig:patterns2}
\end{figure*}
\begin{figure*}
\centering
\includegraphics[width=0.9\linewidth]{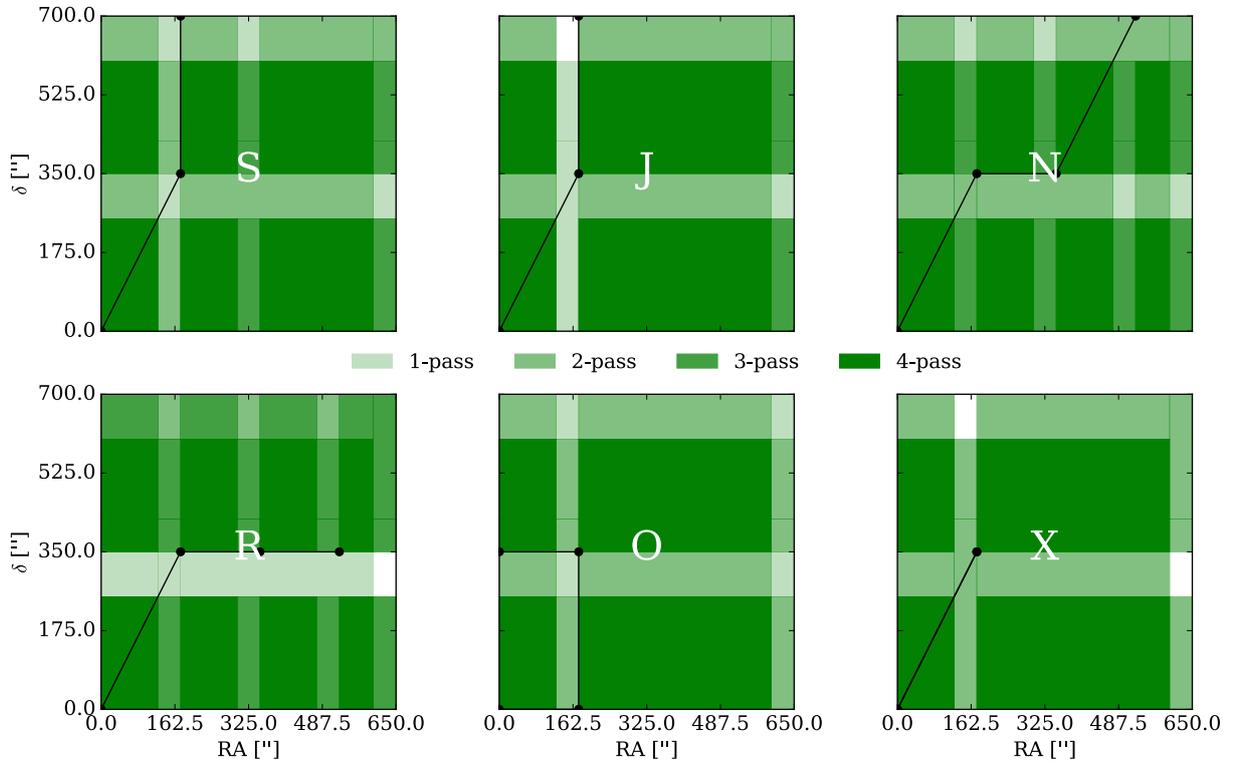}
\caption{The $n$-pass overlap tiles of the S, J, N, R, O, and X patterns from left to right and top to bottom, but with a dither scaling of $r=3.5$, corresponding to the dither step where the first dither aligns perfectly with the vertically adjacent row of detectors, so that the 0-pass coverage is large.}
\label{fig:patterns3}
\end{figure*}



The connection to the dips in calibration quality can be seen particularly well in the visible increase in 0-pass coverage at $d_x = 356\arcsec$.


\begin{figure*}
\centering
\includegraphics[width=\linewidth]{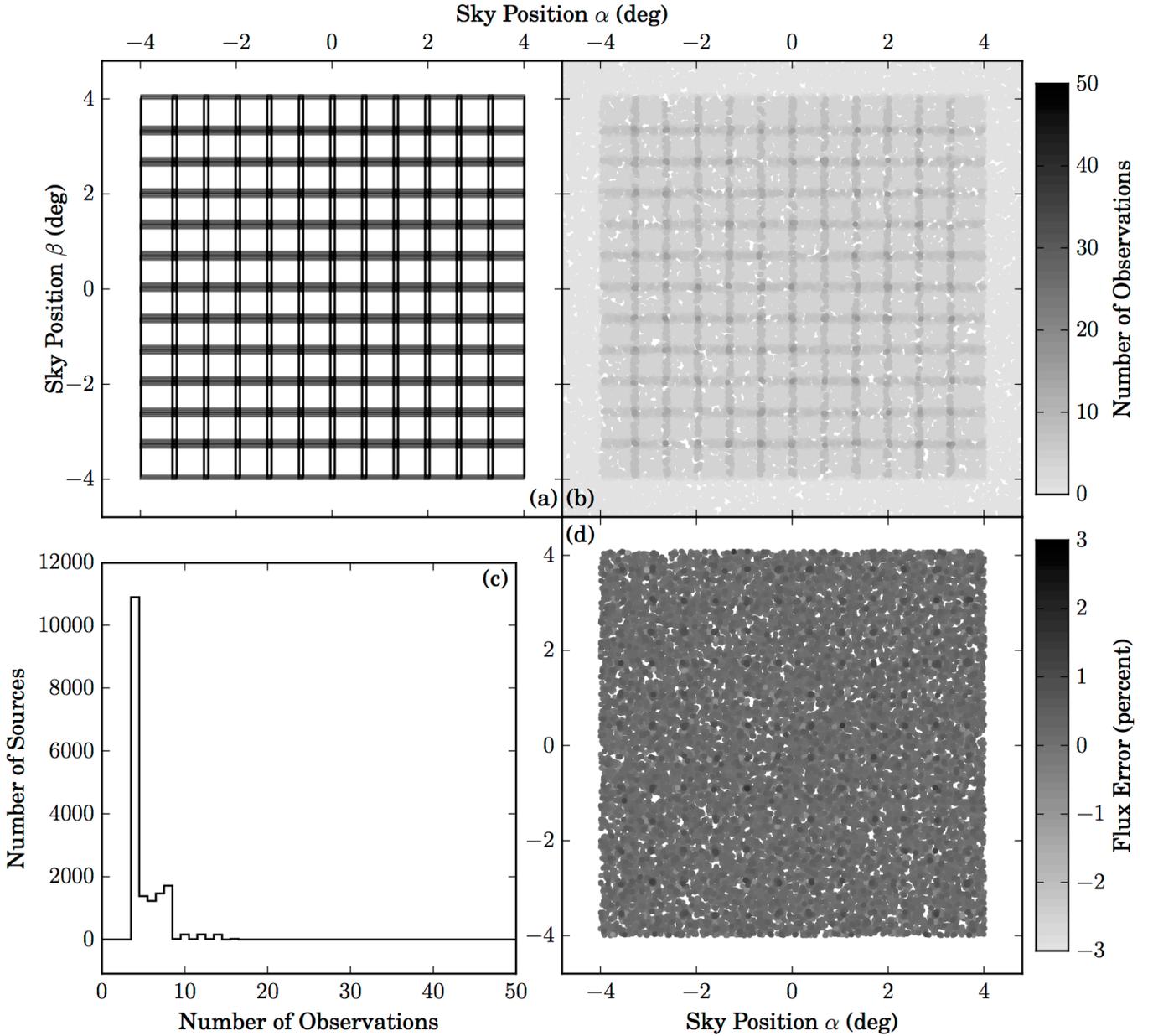}
\caption{J survey configuration, coverage and distribution of calibrators. These plots were generated using the code distributed with \citet{Holmes:2012}, where a) is the survey footprint in the configuration with no detector or pointing gaps, b) is its coverage, c) is the number of observations of sources, and d) is the uniformity of the survey aftera \citet{Holmes:2012}-like \textit{ubercalibration}.}
\label{fig:hoggJ}
\end{figure*}
\section{Fitting the Flat-field}\label{app:FF}

We use the publicly available \href{https://github.com/davidwhogg/SelfCalibration}{\textit{Self-Calibration}} code to repeat the analysis of \citet{Holmes:2012} using two (J and S) out of the above 8 dither patterns and compare them to two of the \citet{Holmes:2012} dither patterns (uniformly displaced, C and quasi-random, QR). This analysis considers the very small scales of calibration and tries to fit the pixel-to-pixel flat-field residuals across the full field of view, which is modelled as a square and not split into detector squares. Therefore there are no gaps between detectors or pointings and the coverage is always at least 4-pass. We re-plot some of the figures from their paper with these four dither patterns and with only 4 dithers per pointing (they had 9) and show the results in \autoref{tab:ff}, \autoref{fig:hoggS}, \autoref{fig:hoggJ}, \autoref{fig:FF-S} and \autoref{fig:FF-Qr}. We do this to compare the new pattern we propose, S, to the previously proposed patterns.

\begin{table}
\begin{tabularx}{\linewidth}{|X|r|r|r|r|r|}
\hline
pat. & \# iter. & RMS & Badness & BiB & $\chi^2$ \\
\hline\hline
C & 26.0 & 0.2098 & 0.3373 & 0.0233 & 315496.97 \\
QR & 13.0 & 0.2055 & 0.3388 & 0.0059 & 36554.80 \\
J & 3459.0 & 0.2538 & 0.3340 & 0.0825 & 307420.63 \\
S & 1447.0 & 0.2491 & 0.3364 & 0.0764 & 307616.41 \\
\hline
\end{tabularx}
\caption{Results of the \citet{Holmes:2012}-like analysis of the S and J patterns. The columns show the number of iterations, the source RMS, the Badness metric, the Best-in-Basis metric and the fit $\chi^2$ for each of the 4 patterns.}
\label{tab:ff}
\end{table}

\begin{figure*}
\centering
\includegraphics[width=\linewidth]{\plotpath{S-survey}}
\caption{S survey configuration, coverage and distribution of calibrators.These plots were generated using the code distributed with \citet{Holmes:2012}, where a) is the survey footprint in the configuration with no detector or pointing gaps, b) is its coverage, c) is the number of observations of sources, and d) is the uniformity of the survey aftera \citet{Holmes:2012}-like \textit{ubercalibration}.}
\label{fig:hoggS}
\end{figure*}

\begin{figure*}
\centering
\includegraphics[width=\linewidth]{\plotpath{S-FF}}
\caption{S-pattern flat-field fits. These plots were generated using the code distributed with \citet{Holmes:2012}. Here, a) and c) are the comparisons of the achieved fit and the best possiblt fit (so-called ``best-in-basis'') in grey, and the true flat-field model in black. On the right are the b) and d) plots, which show the residuals compared to the true flat-field for the achieved and best-in-basis fits respectively.}
\label{fig:FF-S}
\end{figure*}

\begin{figure*}
\centering
\includegraphics[width=\linewidth]{\plotpath{Qr-FF}}
\caption{Quasi-random flat-field fits. These plots were generated using the code distributed with \citet{Holmes:2012}. Here, a) and c) are the comparisons of the achieved fit and the best possiblt fit (so-called ``best-in-basis'') in grey, and the true flat-field model in black. On the right are the b) and d) plots, which show the residuals compared to the true flat-field for the achieved and best-in-basis fits respectively.}
\label{fig:FF-Qr}
\end{figure*}


\begin{thebibliography}{}
\makeatletter
\relax
\def\mn@urlcharsother{\let\do\@makeother \do\$\do\&\do\#\do\^\do\_\do\%\do\~}
\def\mn@doi{\begingroup\mn@urlcharsother \@ifnextchar [ {\mn@doi@}
  {\mn@doi@[]}}
\def\mn@doi@[#1]#2{\def\@tempa{#1}\ifx\@tempa\@empty \href
  {http://dx.doi.org/#2} {doi:#2}\else \href {http://dx.doi.org/#2} {#1}\fi
  \endgroup}
\def\mn@eprint#1#2{\mn@eprint@#1:#2::\@nil}
\def\mn@eprint@arXiv#1{\href {http://arxiv.org/abs/#1} {{\tt arXiv:#1}}}
\def\mn@eprint@dblp#1{\href {http://dblp.uni-trier.de/rec/bibtex/#1.xml}
  {dblp:#1}}
\def\mn@eprint@#1:#2:#3:#4\@nil{\def\@tempa {#1}\def\@tempb {#2}\def\@tempc
  {#3}\ifx \@tempc \@empty \let \@tempc \@tempb \let \@tempb \@tempa \fi \ifx
  \@tempb \@empty \def\@tempb {arXiv}\fi \@ifundefined
  {mn@eprint@\@tempb}{\@tempb:\@tempc}{\expandafter \expandafter \csname
  mn@eprint@\@tempb\endcsname \expandafter{\@tempc}}}

\bibitem[\protect\citeauthoryear{{Agarwal} et~al.,}{{Agarwal}
  et~al.}{2014}]{Agarwal:2014}
{Agarwal} N.,  et~al., 2014, \mn@doi [\jcap] {10.1088/1475-7516/2014/04/007},
  \href {http://adsabs.harvard.edu/abs/2014JCAP...04..007A} {4, 7}

\bibitem[\protect\citeauthoryear{{Amendola} et~al.,}{{Amendola}
  et~al.}{2013}]{Amendola:2013}
{Amendola} L.,  et~al., 2013, \mn@doi [Living Reviews in Relativity]
  {10.12942/lrr-2013-6}, \href
  {http://adsabs.harvard.edu/abs/2013LRR....16....6A} {16, 6}

\bibitem[\protect\citeauthoryear{{Astropy Collaboration} et~al.,}{{Astropy
  Collaboration} et~al.}{2013}]{Astropy:2013}
{Astropy Collaboration} et~al., 2013, \mn@doi [\aap]
  {10.1051/0004-6361/201322068}, \href
  {http://adsabs.harvard.edu/abs/2013A%26A...558A..33A} {558, A33}

\bibitem[\protect\citeauthoryear{{Awan} et~al.,}{{Awan}
  et~al.}{2016}]{Awan:2016}
{Awan} H.,  et~al., 2016, \mn@doi [\apj] {10.3847/0004-637X/829/1/50}, \href
  {http://adsabs.harvard.edu/abs/2016ApJ...829...50A} {829, 50}

\bibitem[\protect\citeauthoryear{Beletic et~al.,}{Beletic
  et~al.}{2008}]{Beletic:2008}
Beletic J.~W.,  et~al., 2008, \mn@doi [Proc. SPIE] {10.1117/12.790382}, 7021,
  70210H

\bibitem[\protect\citeauthoryear{{Bridle}, {Crittenden}, {Melchiorri},
  {Hobson}, {Kneissl}  \& {Lasenby}}{{Bridle} et~al.}{2002}]{Bridle:2002}
{Bridle} S.~L.,  {Crittenden} R.,  {Melchiorri} A.,  {Hobson} M.~P.,  {Kneissl}
  R.,   {Lasenby} A.~N.,  2002, \mn@doi [\mnras]
  {10.1046/j.1365-8711.2002.05709.x}, \href
  {http://adsabs.harvard.edu/abs/2002MNRAS.335.1193B} {335, 1193}

\bibitem[\protect\citeauthoryear{{Cannon} \& {Pickering}}{{Cannon} \&
  {Pickering}}{1993}]{Cannon:1925}
{Cannon} A.~J.,  {Pickering} E.~C.,  1993, VizieR Online Data Catalog, \href
  {http://adsabs.harvard.edu/abs/1993yCat.3135....0C} {3135}

\bibitem[\protect\citeauthoryear{{Crocce} et~al.,}{{Crocce}
  et~al.}{2016}]{Crocce:2015}
{Crocce} M.,  et~al., 2016, \mn@doi [Mon. Not. Roy. Astron. Soc.]
  {10.1093/mnras/stv2590}, \href
  {http://adsabs.harvard.edu/abs/2015arXiv150705360C} {455, 4301}

\bibitem[\protect\citeauthoryear{{Cropper} et~al.,}{{Cropper}
  et~al.}{2014}]{Cropper:2014}
{Cropper} M.,  et~al., 2014, in Space Telescopes and Instrumentation 2014:
  Optical, Infrared, and Millimeter Wave. p. 91430J,
  \mn@doi{10.1117/12.2055543}

\bibitem[\protect\citeauthoryear{{Dawson} et~al.,}{{Dawson}
  et~al.}{2013}]{Dawson:2013}
{Dawson} K.~S.,  et~al., 2013, \mn@doi [\aj] {10.1088/0004-6256/145/1/10},
  \href {http://adsabs.harvard.edu/abs/2013AJ....145...10D} {145, 10}

\bibitem[\protect\citeauthoryear{{Eisenstein} et~al.,}{{Eisenstein}
  et~al.}{2011}]{Eisenstein:2011}
{Eisenstein} D.~J.,  et~al., 2011, \mn@doi [\aj] {10.1088/0004-6256/142/3/72},
  \href {http://adsabs.harvard.edu/abs/2011AJ....142...72E} {142, 72}

\bibitem[\protect\citeauthoryear{{Elsner}, {Leistedt}  \& {Peiris}}{{Elsner}
  et~al.}{2016}]{Elsner:2015}
{Elsner} F.,  {Leistedt} B.,   {Peiris} H.~V.,  2016, \mn@doi [\mnras]
  {10.1093/mnras/stv2777}, \href
  {http://adsabs.harvard.edu/abs/2016MNRAS.456.2095E} {456, 2095}

\bibitem[\protect\citeauthoryear{{Finkbeiner} et~al.,}{{Finkbeiner}
  et~al.}{2016}]{Finkbeiner:2015}
{Finkbeiner} D.~P.,  et~al., 2016, \mn@doi [\apj] {10.3847/0004-637X/822/2/66},
  \href {http://adsabs.harvard.edu/abs/2016ApJ...822...66F} {822, 66}

\bibitem[\protect\citeauthoryear{{Garilli} et~al.,}{{Garilli}
  et~al.}{2014}]{Garilli:2014}
{Garilli} B.,  et~al., 2014, \mn@doi [\aap] {10.1051/0004-6361/201322790},
  \href {http://adsabs.harvard.edu/abs/2014A%26A...562A..23G} {562, A23}

\bibitem[\protect\citeauthoryear{{Garilli} et~al.}{{Garilli}
  et~al.}{prep}]{Imodel}
{Garilli} B.,  et~al., in prep.

\bibitem[\protect\citeauthoryear{{Girardi}, {Groenewegen}, {Hatziminaoglou}  \&
  {da Costa}}{{Girardi} et~al.}{2005}]{Girardi:2005}
{Girardi} L.,  {Groenewegen} M.~A.~T.,  {Hatziminaoglou} E.,   {da Costa} L.,
  2005, \mn@doi [\aap] {10.1051/0004-6361:20042352}, \href
  {http://cdsads.u-strasbg.fr/abs/2005A%26A...436..895G} {436, 895}

\bibitem[\protect\citeauthoryear{{Guzzo} et~al.,}{{Guzzo}
  et~al.}{2014}]{Guzzo:2014}
{Guzzo} L.,  et~al., 2014, \mn@doi [\aap] {10.1051/0004-6361/201321489}, \href
  {http://adsabs.harvard.edu/abs/2014A%26A...566A.108G} {566, A108}

\bibitem[\protect\citeauthoryear{{Ho} et~al.,}{{Ho} et~al.}{2012}]{Ho:2012}
{Ho} S.,  et~al., 2012, \mn@doi [\apj] {10.1088/0004-637X/761/1/14}, \href
  {http://adsabs.harvard.edu/abs/2012ApJ...761...14H} {761, 14}

\bibitem[\protect\citeauthoryear{{Holmes}, {Hogg}  \& {Rix}}{{Holmes}
  et~al.}{2012}]{Holmes:2012}
{Holmes} R.,  {Hogg} D.~W.,   {Rix} H.-W.,  2012, \mn@doi [\pasp]
  {10.1086/668656}, \href {http://adsabs.harvard.edu/abs/2012PASP..124.1219H}
  {124, 1219}

\bibitem[\protect\citeauthoryear{{Kaiser} et~al.,}{{Kaiser}
  et~al.}{2010}]{Kaiser:2010}
{Kaiser} N.,  et~al., 2010, in Ground-based and Airborne Telescopes III. p.
  77330E, \mn@doi{10.1117/12.859188}

\bibitem[\protect\citeauthoryear{{Laureijs} et~al.,}{{Laureijs}
  et~al.}{2011}]{Laureijs:2011}
{Laureijs} R.,  et~al., 2011, ESA report, \href
  {http://adsabs.harvard.edu/abs/2011arXiv1110.3193L} {ESA/SRE(2011)12}

\bibitem[\protect\citeauthoryear{{Leistedt} et~al.,}{{Leistedt}
  et~al.}{2016}]{Leistedt:2015}
{Leistedt} B.,  et~al., 2016, \mn@doi [\apjs] {10.3847/0067-0049/226/2/24},
  \href {http://adsabs.harvard.edu/abs/2016ApJS..226...24L} {226, 24}

\bibitem[\protect\citeauthoryear{{Maciaszek} et~al.,}{{Maciaszek}
  et~al.}{2014}]{Maciaszek:2014}
{Maciaszek} T.,  et~al., 2014, in Space Telescopes and Instrumentation 2014:
  Optical, Infrared, and Millimeter Wave. p. 91430K,
  \mn@doi{10.1117/12.2056702}

\bibitem[\protect\citeauthoryear{{Maiorano} et~al.}{{Maiorano}
  et~al.}{prep}]{Maiorano:2017}
{Maiorano} E.,  et~al., in prep.

\bibitem[\protect\citeauthoryear{{Oke} \& {Gunn}}{{Oke} \&
  {Gunn}}{1983}]{Oke:1983}
{Oke} J.~B.,  {Gunn} J.~E.,  1983, \mn@doi [\apj] {10.1086/160817}, \href
  {http://adsabs.harvard.edu/abs/1983ApJ...266..713O} {266, 713}

\bibitem[\protect\citeauthoryear{{Padmanabhan} et~al.,}{{Padmanabhan}
  et~al.}{2008}]{Padmanabhan:2008}
{Padmanabhan} N.,  et~al., 2008, \mn@doi [\apj] {10.1086/524677}, \href
  {http://adsabs.harvard.edu/abs/2008ApJ...674.1217P} {674, 1217}

\bibitem[\protect\citeauthoryear{{Pickles}}{{Pickles}}{1998}]{Pickles:1998}
{Pickles} A.~J.,  1998, \mn@doi [\pasp] {10.1086/316197}, \href
  {http://adsabs.harvard.edu/abs/1998PASP..110..863P} {110, 863}

\bibitem[\protect\citeauthoryear{{Ross} et~al.,}{{Ross}
  et~al.}{2011}]{Ross:2011}
{Ross} A.~J.,  et~al., 2011, \mn@doi [\mnras]
  {10.1111/j.1365-2966.2011.19351.x}, \href
  {http://adsabs.harvard.edu/abs/2011MNRAS.417.1350R} {417, 1350}

\bibitem[\protect\citeauthoryear{{Ross} et~al.,}{{Ross}
  et~al.}{2012}]{Ross:2012}
{Ross} A.~J.,  et~al., 2012, \mn@doi [\mnras]
  {10.1111/j.1365-2966.2012.21235.x}, \href
  {http://adsabs.harvard.edu/abs/2012MNRAS.424..564R} {424, 564}

\bibitem[\protect\citeauthoryear{{Scaramella} et~al.}{{Scaramella}
  et~al.}{prep}]{Scaramella:2017}
{Scaramella} R.,  et~al., in prep.

\bibitem[\protect\citeauthoryear{{Schlafly} et~al.,}{{Schlafly}
  et~al.}{2012}]{Schlafly:2012}
{Schlafly} E.~F.,  et~al., 2012, \mn@doi [\apj] {10.1088/0004-637X/756/2/158},
  \href {http://adsabs.harvard.edu/abs/2012ApJ...756..158S} {756, 158}

\bibitem[\protect\citeauthoryear{{Shafer} \& {Huterer}}{{Shafer} \&
  {Huterer}}{2015}]{Shafer:2015}
{Shafer} D.~L.,  {Huterer} D.,  2015, \mn@doi [\mnras] {10.1093/mnras/stu2640},
  \href {http://adsabs.harvard.edu/abs/2015MNRAS.447.2961S} {447, 2961}

\bibitem[\protect\citeauthoryear{{Stubbs} \& {Tonry}}{{Stubbs} \&
  {Tonry}}{2006}]{Stubbs:2006}
{Stubbs} C.~W.,  {Tonry} J.~L.,  2006, \mn@doi [\apj] {10.1086/505138}, \href
  {http://adsabs.harvard.edu/abs/2006ApJ...646.1436S} {646, 1436}

\bibitem[\protect\citeauthoryear{{Swanson}, {Tegmark}, {Hamilton}  \&
  {Hill}}{{Swanson} et~al.}{2008}]{Swanson:2008}
{Swanson} M.~E.~C.,  {Tegmark} M.,  {Hamilton} A.~J.~S.,   {Hill} J.~C.,  2008,
  \mn@doi [\mnras] {10.1111/j.1365-2966.2008.13296.x}, \href
  {http://adsabs.harvard.edu/abs/2008MNRAS.387.1391S} {387, 1391}

\bibitem[\protect\citeauthoryear{{Vanhollebeke}, {Groenewegen}  \&
  {Girardi}}{{Vanhollebeke} et~al.}{2009}]{Vanhollebeke:2009}
{Vanhollebeke} E.,  {Groenewegen} M.~A.~T.,   {Girardi} L.,  2009, \mn@doi
  [\aap] {10.1051/0004-6361/20078472}, \href
  {http://adsabs.harvard.edu/abs/2009A%26A...498...95V} {498, 95}

\bibitem[\protect\citeauthoryear{{Vargas Maga{\~n}a} et~al.,}{{Vargas
  Maga{\~n}a} et~al.}{2013}]{Vargas:2013}
{Vargas Maga{\~n}a} M.,  et~al., 2013, preprint, \href
  {http://adsabs.harvard.edu/abs/2013arXiv1312.4996V} {} (\mn@eprint {arXiv}
  {1312.4996})

\bibitem[\protect\citeauthoryear{{Zoubian} et~al.,}{{Zoubian}
  et~al.}{2014}]{Zoubian:2014}
{Zoubian} J.,  et~al., 2014, in {Manset} N.,  {Forshay} P.,  eds,  Astronomical
  Society of the Pacific Conference Series Vol. 485, Astronomical Data Analysis
  Software and Systems XXIII. p.~509

\makeatother
\end{thebibliography}
\end{document}